%% file: main.tex
\documentclass[conference]{IEEEtran}

\usepackage{amsmath}
\usepackage{amssymb}
\usepackage{amsthm}
\usepackage{array}
\usepackage{booktabs}
\usepackage{graphicx}
\usepackage{cite}
\usepackage{url}
\usepackage{pifont}
\usepackage{algorithm}
\usepackage[noend]{algpseudocode}
\usepackage{enumitem}
\usepackage[table]{xcolor}
\usepackage{tikz}
\usetikzlibrary{arrows.meta,positioning,calc,fit,backgrounds}

\definecolor{cC}{RGB}{31,78,140}
\definecolor{cF}{RGB}{192,57,43}

\theoremstyle{plain}
\newtheorem{theorem}{Theorem}
\newtheorem{lemma}{Lemma}

\newtheorem{proposition}{Proposition}
\theoremstyle{definition}
\newtheorem{definition}{Definition}
\newtheorem{problem}{Problem}
\theoremstyle{remark}

\newcommand{\ravel}{\textsc{Virel}}
\newcommand{\vex}{\textsc{Virel}-Exact}
\newcommand{\veb}{\textsc{Virel}-EB}
\newcommand{\fast}{\textsc{Fast}}
\newcommand{\upper}{\textsc{Upper}}
\newcommand{\bits}{\operatorname{bits}}
\newcommand{\rint}{\operatorname{rint}}
\newcommand{\fl}{\operatorname{fl}}
\newcommand{\zz}{\operatorname{zz}}
\newcommand{\dbl}{\operatorname{double}}

\title{VIREL: Route-Local Lattice Residual\\
Compression for Exact and Error-Bounded\\
Floating-Point Time Series}

\author{
\IEEEauthorblockN{Yue Zhang}
\IEEEauthorblockA{Shanghai Jiao Tong University\\
Shanghai, China\\
zhangyue20040611@sjtu.edu.cn}
\and
\IEEEauthorblockN{Jiatao Lin}
\IEEEauthorblockA{Shanghai Jiao Tong University\\
Shanghai, China\\
1169652042@sjtu.edu.cn}
\and
\IEEEauthorblockN{Haopeng Chen}
\IEEEauthorblockA{Shanghai Jiao Tong University\\
Shanghai, China\\
chen-hp@sjtu.edu.cn}
}

\begin{document}
\maketitle

\begin{figure}[!b]
\centering
\includegraphics[width=\columnwidth]{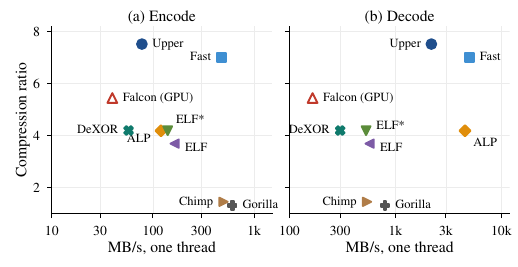}
\caption{Compression ratio against one-thread throughput on the 14 canonical
streams with independent 1,024-value pages.  Both \vex{} profiles lie above
every baseline, and \fast{} also outruns ALP in both directions.}
\label{fig:tradeoff}
\end{figure}

\begin{abstract}
Time-series databases store floating-point columns in independent pages that
are written at ingest, rewritten at compaction, and decoded on every scan.
Some columns must return the stored IEEE~754 word; others accept a pointwise
bound~$\varepsilon$.  Codecs for both contracts turn each value into an
integer on a lattice, but a real page mixes decimal resolutions.  Keeping one
scale per block charges every coarse step for the finest value: one scale-6
reading in a 1,024-value page cuts Falcon's compression ratio by 36\%.
We present \ravel{}, route-local lattice residual compression.  A value
enters a lattice only after the encoder has replayed the decoder's arithmetic
on it.  \vex{} replays the decoder's division at each value's least decimal
scale, which we prove always succeeds below $2^{51}$, gives every resolution
its own lane and history, and places averages on a rational lattice; the same
finer value costs it 0.6\%.  \veb{} replays a fused multiply-add on an error
lattice, so every admitted value decodes within $\varepsilon$.  One integer
coder serves both contracts.
On 14 canonical streams, \vex{} emits 27.6\% fewer bytes than Falcon.  \fast{}
beats ALP on ratio and both speeds at once: 7.01$\times$ against 4.19$\times$,
476 against 120~MB/s, and 5.0 against 4.5~GB/s on one core.  \upper{} is the
smallest output on all 48 larger streams.  Ingest with \fast{} and
compaction with \upper{} store those streams in Apache TsFile at 48.7\% of
ALP's bytes.  \veb{} beats Machete on ratio and
both speeds in all 90 dataset--bound cells.
\end{abstract}


\input{sections/01_introduction}
\input{sections/02_background}
\input{sections/03_overview}
\input{sections/04_exact}
\input{sections/05_bounded}
\input{sections/06_coder}
\input{sections/07_evaluation}
\input{sections/08_related_work}
\input{sections/09_conclusion}

\bibliographystyle{IEEEtran}
\bibliography{references}

\end{document}

%% file: sections/01_introduction.tex
\section{Introduction}
\label{sec:introduction}

Sensing devices now monitor systems from a household appliance to a
distributed industrial plant, and each deployment records its readings as time
series \cite{tsdbsurvey}.  The sensors on a Boeing~787 produce upwards of
half a terabyte of data on one flight \cite{tsdbsurvey}.  The series are
retained and scanned, so the same values are written at ingest, sometimes
rewritten at compaction, and decoded on every query.  In our TsFile workload
the measurements are 97.0\% of the file once timestamps, headers, and indexes
are counted.  The codec on that column sets the disk footprint, the I/O of
every scan, and how much history fits in cache.

Compressing the column is how a time-series store pays for this volume, and
the contract travels with the column.  Audit logs, financial ticks, and
regulated records must return the stored IEEE~754 words \cite{ieee754},
including signed zeros and NaN payloads \cite{falcon}.  Monitoring and
analytics columns accept a pointwise bound: each decoded value lies within
$\varepsilon$ of the stored one \cite{machete,serf,sz}.  Production systems
already cut each column into independent pages.  Apache IoTDB writes TsFile
chunks of fixed-size pages \cite{iotdb,tsfile}, Gorilla keeps in-memory blocks
\cite{gorilla}, and log-structured engines write a page once at ingest and
rewrite it during compaction \cite{lsm}.  The page bounds encoder memory,
isolates a corrupt page, and lets a scan skip by header statistics.

Separate codec families serve the two contracts on these pages.  Decimal
codecs such as ALP, DeXOR, and Falcon turn readings into integers at a
decimal scale \cite{alp,dexor,falcon}; quantizing codecs such as SZ, Machete,
and Serf turn them into integers on a grid of step $2\varepsilon$
\cite{sz,machete,serf}.  Both families code integers on a lattice, and both
rely on the decoder to map each integer back to a double with one
floating-point operation, a division by $10^{s}$ or a multiply-add with the
grid step.

\begin{figure*}[t]
  \centering
  \resizebox{\textwidth}{!}{\input{figures/fig_strategies}}
  \caption{One page, three ways to hold a second resolution; bit counts
  exclude the first value.  (a)~Lifting every value to the finest scale
  multiplies the coarse steps by $10^{2}$.  (b)~Keeping the coarse scale turns
  each fine value into a raw word.  (c)~\vex{} gives each resolution a lane
  with its own history, and an exact gcd shrinks the coarse steps further.}
  \label{fig:strategies}
\end{figure*}

\textbf{Challenge I: faithful coordinates under binary64.}  The encoder
computes $x\times10^{s}$ or $x/2\varepsilon$ in binary64, the decoder inverts
it in binary64, and either side can round.  Closed-form rules certify the
round trip under conditions such as ``at most 15 significant digits''
\cite{falcon}, so values outside a rule leave the integer path even when
their coordinate would survive the round trip.  Negative zero is the smallest
case: its integer is 0, and $0/10^{s}$ returns $+0.0$.

\textbf{Challenge II: mixed resolutions inside a page.}  A thermometer reports
12.579, a calibration step writes 12.59621, and a unit conversion appends
further digits.  On the canonical streams, 97.3\% of the pages hold values of
several minimal decimal scales, and one page in five holds values finer than
the scale most of its values share.  A one-scale codec either lifts every value to the
finest scale, which turns coarse steps such as $+10,+5$ into $+1000,+500$, or
keeps the coarse scale and stores the misfits as exceptions
(Figure~\ref{fig:strategies}).  Both are expensive.  One scale-6 value per
1,024-value page drops Falcon from 10.31$\times$ to 6.62$\times$, and at 50\%
fine values ALP falls from 6.58$\times$ to 3.10$\times$.  Averaged readings
are harder still: a mean of three tenths prints with 16 digits and lies on no
short decimal lattice at all.

\textbf{Challenge III: two encoders and one decoder at page granularity.}
Pages are small and independent, so every model must be learned and paid for
inside the page.  A page is also encoded twice in its life: at ingest, which
needs hundreds of megabytes per second per core, and at compaction, which can
trade CPU for bytes \cite{rocksdbcompression}, while every scan pays the
decoder.  Existing codecs couple one integer coder to one front end, and a
byte compressor stacked on top adds decode work to every scan and hides the
value structure from the encoder.

We present \ravel{}, route-local lattice residual compression.  \vex{}
replays the decoder's division at each scale and keeps the least scale that
returns the same IEEE word; \veb{} replays a fused multiply-add and keeps a
value only when the result lies within $\varepsilon$.  A value that passes this replay already meets its contract, and the theorems only bound how often the check fails.  An exact page gives each resolution
its own lane and history, at most four lanes, moves averages onto a rational
lattice, and leaves the remaining words on a binary lane.  Both codecs then
use one stream coder: an exact gcd, a coarse/residue split, and the cheapest
of four page-local representations.  \fast{} writes at ingest and skips
entropy decoding; \upper{} rewrites at compaction and prices every candidate
by the bits it emits.

\begin{enumerate}
  \item[(1)] Replay-checked admission for both contracts, with a closed-form
  condition under which decimal replay always succeeds ($|q|<2^{51}$), a
  lifting theorem, and an unconditional error guarantee for the bounded
  lattice (Sections~\ref{sec:exact} and~\ref{sec:bounded}).
  \item[(2)] Route-local lattice residuals for mixed-resolution exact pages, with a lower
  bound on the cost of lifting, rational lattices for averaged readings, and
  route-persistent histories (Section~\ref{sec:exact}).
  \item[(3)] One integer stream coder for exact and bounded pages, with exact
  gcd factoring, a coarse/residue split, four page-local representations,
  vectorized decoding, and two encoder profiles over one format
  (Section~\ref{sec:coder}).
  \item[(4)] An evaluation on 14 canonical streams, 48 larger streams with
  74.7~million values, Apache TsFile storage and compaction, a
  mixed-resolution sweep, and 90 dataset--bound cells for the bounded
  contract, with ablations, scaling, and 136,000 randomized round trips
  (Section~\ref{sec:evaluation}).
\end{enumerate}

Figure~\ref{fig:tradeoff} previews the exact results.  \vex{} \upper{} emits
27.6\% fewer bytes than Falcon, the strongest exact baseline, and is the
smallest exact output on 13 of the 14 canonical streams.  \fast{} emits
40.2\% fewer bytes than ALP while encoding 4.0$\times$ and decoding
1.10$\times$ as fast on one core.  On the 48 larger streams \upper{} is the smallest output on every
stream, and inside Apache TsFile, \fast{} ingest followed by \upper{}
compaction stores the suite in 48.7\% of ALP's bytes.  Under the bounded
contract \veb{} beats Machete on ratio, encode speed, and decode speed in
every one of the 90 dataset--bound cells, by geometric means of 1.56$\times$,
5.18$\times$, and 2.60$\times$.

%% file: figures/fig_strategies.tex
% Column stacks in the Falcon sheet style: offset gray pages, block arrows,
% red only on the fine rows. No enclosing frame.
\begin{tikzpicture}[x=1cm,y=1cm,font=\footnotesize,
  hd/.style={font=\footnotesize\bfseries,inner sep=0pt,anchor=south},
  sub/.style={font=\scriptsize,inner sep=0pt,anchor=south,text=black!60},
  num/.style={inner sep=0pt,anchor=center,font=\footnotesize},
  stp/.style={inner sep=0pt,anchor=west,font=\scriptsize},
  cost/.style={inner sep=0pt,anchor=north,font=\footnotesize\bfseries}]
  \def\ch{0.38}
  \def\sh{2.28}
  \def\yt{0}
  \def\px{0.08}\def\pw{2.28}
  \def\ax{3.55}\def\aw{2.62}
  \def\bx{8.55}\def\bw{2.18}
  \def\cx{12.55}\def\cw{1.78}
  \def\fx{14.91}\def\fw{1.78}
  % offset sheets, then the ruled front page
  \foreach \x/\w in {\px/\pw,\ax/\aw,\bx/\bw,\cx/\cw,\fx/\fw}{
    \fill[black!10] ({\x-0.10},{\yt+0.08}) rectangle ({\x+\w-0.10},{\yt-\sh+0.08});
    \draw[black!35,line width=0.3pt] ({\x-0.10},{\yt+0.08}) rectangle ({\x+\w-0.10},{\yt-\sh+0.08});
    \fill[black!6] ({\x-0.05},{\yt+0.04}) rectangle ({\x+\w-0.05},{\yt-\sh+0.04});
    \draw[black!40,line width=0.3pt] ({\x-0.05},{\yt+0.04}) rectangle ({\x+\w-0.05},{\yt-\sh+0.04});
    \fill[black!3] (\x,\yt) rectangle ({\x+\w},\yt-\sh);
    \draw[black!65,line width=0.45pt] (\x,\yt) rectangle ({\x+\w},\yt-\sh);
    \foreach \k in {1,...,5}
      \draw[black!28,line width=0.25pt] (\x,{-\k*\ch}) -- ++(\w,0);
  }
  % page
  \foreach \i/\v in {0/12.579,1/12.589,2/12.594,3/12.59621,4/12.599,5/12.60381}
    \node[num] at ({\px+0.5*\pw},{-(\i+0.5)*\ch}) {\v};
  % (a) two extra digits stay inside the token
  \foreach \i/\h in {0/12579,1/12589,2/12594,4/12599}
    \node[num] at ({\ax+0.5*\aw},{-(\i+0.5)*\ch}) {\h\textcolor{cF}{00}};
  \foreach \i/\v in {3/1259621,5/1260381}
    \node[num] at ({\ax+0.5*\aw},{-(\i+0.5)*\ch}) {\v};
  \foreach \i/\s in {1/+1000,2/+500,3/+221,4/+279,5/+481}
    \node[stp,text=cF] at ({\ax+\aw+0.08},{-(\i+0.5)*\ch}) {$\s$};
  % (b)
  \foreach \i/\v in {0/12579,1/12589,2/12594,4/12599}
    \node[num] at ({\bx+0.5*\bw},{-(\i+0.5)*\ch}) {\v};
  \node[num] at ({\bx+0.5*\bw},{-(3.5)*\ch}) {raw word};
  \node[num] at ({\bx+0.5*\bw},{-(5.5)*\ch}) {raw word};
  \node[stp,text=black!65] at ({\bx+\bw+0.08},{-(1.5)*\ch}) {$+10$};
  \node[stp,text=black!65] at ({\bx+\bw+0.08},{-(2.5)*\ch}) {$+5$};
  \node[stp,text=black!65] at ({\bx+\bw+0.08},{-(4.5)*\ch}) {$+5$};
  % (c) same six rows
  \foreach \i/\v in {0/12579,1/12589,2/12594,4/12599}
    \node[num] at ({\cx+0.5*\cw},{-(\i+0.5)*\ch}) {\v};
  \node[stp,text=black!75] at ({\cx+\cw+0.07},{-(1.5)*\ch}) {$+2$};
  \node[stp,text=black!75] at ({\cx+\cw+0.07},{-(2.5)*\ch}) {$+1$};
  \node[stp,text=black!75] at ({\cx+\cw+0.07},{-(4.5)*\ch}) {$+1$};
  \node[num] at ({\fx+0.5*\fw},{-(3.5)*\ch}) {1259621};
  \node[num] at ({\fx+0.5*\fw},{-(5.5)*\ch}) {1260381};
  \node[stp,text=black!70] at ({\fx+\fw+0.08},{-(5.5)*\ch}) {$+760$};
  % red stroke on the two fine rows of page, lift, and raw words
  \foreach \x/\w/\i in {\px/\pw/3,\px/\pw/5,\ax/\aw/3,\ax/\aw/5,\bx/\bw/3,\bx/\bw/5}{
    \draw[cF,line width=0.75pt] (\x,{-\i*\ch}) rectangle ++(\w,-\ch);
  }
  % Falcon block arrows in the three gaps
  \foreach \x/\tip in {2.48/3.28,7.42/8.28,11.42/12.28}{
    \fill[black!28] (\x,-0.98) rectangle ({\tip-0.28},-1.30);
    \fill[black!40] ({\tip-0.28},-0.80) -- (\tip,-1.14) -- ({\tip-0.28},-1.48) -- cycle;
  }
  % headers
  \node[hd] at ({\px+0.5*\pw},0.50) {Page};
  \node[hd] at ({\ax+0.5*\aw},0.50) {(a) Lift};
  \node[sub] at ({\ax+0.5*\aw},0.18) {$\times 10^{5}$};
  \node[hd] at ({\bx+0.5*\bw},0.50) {(b) Raw words};
  \node[sub] at ({\bx+0.5*\bw},0.18) {$\times 10^{3}$};
  \node[hd] at ({(\cx+\fx+\fw)/2},0.50) {(c) Lanes};
  \node[sub] at ({\cx+0.5*\cw},0.18) {$s=3$, $\div\,5$};
  \node[sub] at ({\fx+0.5*\fw},0.18) {$s=5$};
  \node[cost,text=cF] at ({\ax+0.5*\aw},-2.38) {45 bits};
  \node[cost,text=cF] at ({\bx+0.5*\bw},-2.38) {170 bits};
  \node[cost] at ({(\cx+\fx+\fw)/2},-2.38) {20 bits};
\end{tikzpicture}

%% file: sections/02_background.tex
\section{Background and Problem}
\label{sec:background}

\textbf{Binary64 arithmetic.}  A binary64 value has a sign bit, an 11-bit
exponent, and a 52-bit fraction with a hidden leading one \cite{ieee754}.  We
write $\bits(x)$ for its 64-bit word, so $+0.0$ and $-0.0$ differ and every
NaN payload is a distinct word.  $\fl(y)$ rounds a real $y$ to the nearest
binary64 value, ties to even.
Products, quotients, and the fused multiply-add are correctly rounded:
$x\otimes y=\fl(xy)$, $x\oslash y=\fl(x/y)$, and
$\mathrm{fma}(a,b,c)=\fl(ab+c)$.  In the normal range
$\fl(y)=y(1+\theta)$ with $|\theta|\le2^{-53}$.  Every integer $N$ with
$|N|\le2^{53}$ is a binary64 value, and so is $10^{s}$ for $s\le22$
\cite{falcon}.  $\rint$ rounds to the nearest integer, and $\fl$ is monotone:
$y\le y'$ implies $\fl(y)\le\fl(y')$.

\textbf{Pages and contracts.}  A page is a sequence $X=\langle
x_0,\ldots,x_{n-1}\rangle$ of $n\le2048$ binary64 values; our storage
experiments use $n=1024$.  A page codec $(\mathcal E,\mathcal D)$ is
page-local: $\mathcal E$ reads only $X$ and $\mathcal D$ reads only
$\mathcal E(X)$, which bounds encoder memory, isolates corruption, and lets a
reader skip a page by its header statistics.

\begin{definition}[Contracts]
\label{def:contracts}
Let $\hat X=\mathcal D(\mathcal E(X))$.  The codec is \emph{exact} if
$\bits(\hat x_i)=\bits(x_i)$ for every $i$.  It is
\emph{$\varepsilon$-bounded} for a binary64 value $\varepsilon>0$ if
$|\hat x_i-x_i|\le\varepsilon$ for every finite $x_i$ and
$\bits(\hat x_i)=\bits(x_i)$ for every other $x_i$.
\end{definition}

\textbf{Lattice coordinates.}  Both contracts are served by mapping values to
integers that the decoder turns back into doubles with one operation.  Under
the exact contract the map is decimal: the \emph{candidate coordinate} of $x$
at scale $s\in\{0,\ldots,18\}$ is $q_s(x)=\rint(x\otimes10^{s})$, and the
decoder's \emph{reconstruction} is $R(q,s)=\dbl(q)\oslash10^{s}$.  Under the
bounded contract the map is a grid of step $\delta$ and origin $o$: the
coordinate is $q=\rint((x-o)/\delta)$ and the reconstruction is
$\mathrm{fma}(q,\delta,o)$.  A page then becomes one or more integer
sequences, and the codec codes their \emph{steps}, the differences of
consecutive coordinates.  We write $\zz(d)=2d$ for $d\ge0$ and $-2d-1$
otherwise, and $\ell(v)=\lfloor\log_2v\rfloor+1$ with $\ell(0)=0$ for the bit
length of $v\ge0$.

\begin{figure}[t]
\centering
\includegraphics[width=\columnwidth]{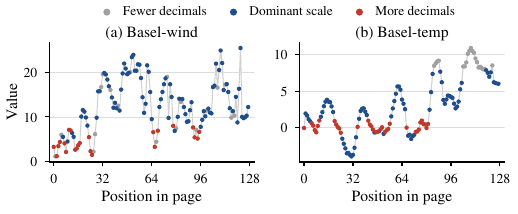}
\caption{Minimal scales on the first 128 values of two canonical pages.  Blue
values sit at the scale that most of the page shares, red ones need more
decimals, and gray ones fewer.}
\label{fig:realpages}
\end{figure}

\textbf{Resolutions in real pages.}  Most extra scales on a page are harmless:
12.5 on a page of readings such as 12.579 has minimal scale~1 but lies on
the page's scale-3 lattice.  The harmful mixes have concrete sources.
Basel-temp and Basel-wind store single-precision readings in their shortest
decimal form, so the number of decimals follows the magnitude
(Figure~\ref{fig:realpages}), and Basel-wind places $1.08$ next to
$3.2399998$, the single-precision products
$0.3\cdot3.6$ and $0.9\cdot3.6$ of a unit conversion.  Stocks-UK quotes in
ticks of 0.1 and 0.05, and averaged sensors land on grids such as
$1/700$ of a unit.  On the canonical streams, values finer than the scale
most of their page shares occur on 19.8\% of the pages and on every page of
the two Basel streams.  There, one scale per page lifts the dominant values
by 1.8 decimal digits on Basel-temp and 2.0 on Basel-wind on average, about
six bits on every step (Section~\ref{sec:lanes}), which is why \vex{} gives
each resolution a lane of its own.

\begin{problem}[Page encoding]
\label{prob:page}
Given a page $X$ and a contract, emit a self-contained page that satisfies the
contract and is small, with a decoder that performs no search and runs in time
linear in $n$, and with two encoders of different cost, one for ingest and one
for compaction, behind one format.
\end{problem}

Problem~\ref{prob:page} has two parts.  The first is admission, deciding which
values may enter a lattice without breaking the contract;
Sections~\ref{sec:exact} and~\ref{sec:bounded} decide it by replay.  The
second is size, keeping the admitted integers small;
Section~\ref{sec:exact} handles mixed resolutions, and
Section~\ref{sec:coder} codes the integer streams of both codecs.

%% file: sections/03_overview.tex
\section{Overview of VIREL}
\label{sec:overview}

\begin{figure*}[t]
  \centering
  \resizebox{\textwidth}{!}{\input{figures/fig_architecture}}
  \caption{\ravel{} architecture.  Two replay-checked front ends
  (\ding{172}--\ding{173}) turn a page into integer sequences, one per lane
  under the exact contract and one per page under the bounded contract, and
  one stream coder (\ding{174}--\ding{176}) writes all of them into a
  self-contained page.  The decoder (bottom) reads every choice from a header
  and performs no search.}
  \label{fig:architecture}
\end{figure*}

Figure~\ref{fig:architecture} shows the two codecs.  Each pairs a front end
that decides lattice membership by replay with a back end that both share.
Section~\ref{sec:lanes} routes the page of Figure~\ref{fig:strategies} to two
lanes, and Section~\ref{sec:coder} factors the gcd of its coarse lane.

\textbf{Front ends.}  \vex{} (Section~\ref{sec:exact}) admits each value at its
minimal decimal scale by replaying $R(q,s)$ (\ding{172}), groups the page's
resolutions into at most four \emph{lanes} by scale interval, and records a
\emph{route}, the lane of every position (\ding{173}).  A decimal lane may move
to a rational lattice when its values are means or interpolations, and a
binary lane carries the words that no decimal lane holds.  Each lane yields
one coordinate sequence with its own history.  \veb{}
(Section~\ref{sec:bounded}) places the whole page on one error lattice, whose
origin it chooses per page, and admits each value by replaying the decoder's
fused multiply-add.  In both codecs a value that fails admission keeps its
IEEE word, so modeling choices change only the size of a page.

\textbf{Shared back end.}  The stream coder (Section~\ref{sec:coder}) takes an
integer sequence, factors the exact gcd of its steps (\ding{174}), optionally
splits a coarse lattice from a residue (\ding{175}), and prices four
page-local representations (\ding{176}): a base value with sparse exceptions,
8-value frame-of-reference blocks, four-way rANS over a literal/escape
alphabet, and four-way rANS over a dictionary.  It writes the cheapest and
records the choice in the stream header, so the decoder never searches.

\textbf{Profiles.}  \vex{} has two encoder profiles over one page format.
\fast{} plans at most three decimal lanes plus the binary lane, ranks lane
sets with a block-width proxy and prices only the winner exactly, codes first
differences, and restricts every stream to the base and block
representations, so its pages decode with shifts, masks, and prefix sums and
no entropy decoding.  \upper{} grows the plan greedily to at most four lanes
and prices every candidate exactly, under every lane variant including second
differences and all four representations; \veb{} also uses all four.  Because
the profiles
differ only in encoder decisions, a storage engine can flush with \fast{} and
let compaction rewrite the same columns with \upper{}, while one decoder
serves all files.

%% file: figures/fig_architecture.tex
% Falcon Fig. 4 grammar, full width: gold slabs, one cyan band, flat stages.
\begin{tikzpicture}[x=1cm,y=1cm,font=\footnotesize,
  >={Stealth[length=4.5pt,width=3.6pt]},
  arr/.style={->,draw=black!70,line width=0.55pt},
  stg/.style={draw=cF!50!black,fill=cF!8,rounded corners=1.2pt,align=center,
              inner xsep=3pt,inner ysep=1.5pt,minimum height=0.64cm,
              font=\footnotesize},
  tag/.style={font=\scriptsize\bfseries,inner sep=0pt,anchor=south west}]
  \newcommand{\sheet}[4]{%
    \foreach \i in {6,5,4,3,2,1}{
      \draw[draw=black!38,fill=yellow!18,line width=0.22pt]
        ({#1-0.048*\i},{#2+0.036*\i}) rectangle ++(#3,#4);
    }
    \draw[draw=black!55,fill=yellow!8,line width=0.4pt]
      (#1,#2) rectangle ++(#3,#4);
  }
  \sheet{0.48}{0.00}{1.22}{2.28}
  \foreach \k/\col in {0/cC,1/cC,2/cC,3/cF,4/cC,5/cF}
    \fill[\col!78] (0.70,{1.78-0.22*\k}) rectangle ++(0.78,0.075);
  \node[font=\footnotesize\bfseries,anchor=south,inner sep=1pt] at (0.95,2.58) {Page};
  \fill[cC!7,rounded corners=1.5pt] (2.05,-0.02) rectangle (14.45,2.48);
  \draw[cC!28,rounded corners=1.5pt] (2.05,-0.02) rectangle (14.45,2.48);
  \node[tag,text=cC!80!black] at (3.20,2.18) {Exact};
  \node[tag,text=cF!85!black] at (3.20,0.78) {Bounded};
  \node[stg,text width=1.68cm] (e1) at (4.15,1.82)
    {\textbf{\ding{172}}\ Replay\\[-1.5pt]{\scriptsize least scale}};
  \node[stg,text width=1.68cm] (e2) at (6.30,1.82)
    {\textbf{\ding{173}}\ Lanes\\[-1.5pt]{\scriptsize route}};
  \node[stg,text width=1.68cm] (b1) at (4.15,0.40)
    {\textbf{\ding{172}}\ Replay\\[-1.5pt]{\scriptsize within $\varepsilon$}};
  \node[stg,text width=1.68cm] (b2) at (6.30,0.40)
    {\textbf{\ding{173}}\ Lattice\\[-1.5pt]{\scriptsize one origin}};
  \node[stg,minimum width=1.40cm] (g) at (8.95,1.11) {\textbf{\ding{174}}\ Gcd};
  \node[stg,minimum width=1.40cm] (sp) at (10.65,1.11) {\textbf{\ding{175}}\ Split};
  \node[stg,minimum width=2.45cm] (cd) at (12.75,1.11)
    {\textbf{\ding{176}}\ Code\\[-1.5pt]{\scriptsize base, blocks, rANS}};
  \sheet{15.20}{0.00}{1.78}{2.28}
  \node[font=\scriptsize,text=black!75] at (16.09,1.62) {header};
  \draw[black!28] (15.42,1.40) -- (16.76,1.40);
  \node[font=\scriptsize,text=cC!80!black] at (16.09,1.16) {route};
  \draw[black!28] (15.42,0.94) -- (16.76,0.94);
  \node[font=\scriptsize] at (16.09,0.68) {streams};
  \node[font=\footnotesize\bfseries,anchor=south,inner sep=1pt] at (15.85,2.58) {Stored page};
  \coordinate (j0) at (2.28,1.11);
  \draw (1.78,1.11) -- (j0);
  \draw[arr] (j0) |- (e1.west);
  \draw[arr] (j0) |- (b1.west);
  \draw[arr] (e1) -- (e2);
  \draw[arr] (b1) -- (b2);
  \coordinate (j1) at ([xshift=-0.28cm]g.west);
  \draw (e2.east) -- (e2.east -| j1) -- (j1);
  \draw (b2.east) -- (b2.east -| j1) -- (j1);
  \draw[arr] (j1) -- (g.west);
  \draw[arr] (g) -- (sp);
  \draw[arr] (sp) -- (cd);
  \draw[arr] (cd.east) -- (14.85,1.11);
  \fill[cC!7,rounded corners=1.5pt] (0.15,-0.48) rectangle (17.40,-1.40);
  \draw[cC!28,rounded corners=1.5pt] (0.15,-0.48) rectangle (17.40,-1.40);
  \node[font=\footnotesize\bfseries,anchor=west] at (0.32,-0.94) {Decode};
  \node[stg,minimum width=2.15cm] (d1) at (3.15,-0.94) {Read header};
  \node[stg,minimum width=1.90cm] (d2) at (6.35,-0.94) {Unpack};
  \node[stg,minimum width=1.90cm] (d3) at (9.45,-0.94) {Restore};
  \node[stg,minimum width=1.90cm] (d4) at (12.55,-0.94) {Replay};
  \node[stg,minimum width=1.90cm] (d5) at (15.70,-0.94) {Scatter};
  \draw[arr] (d1) -- (d2);
  \draw[arr] (d2) -- (d3);
  \draw[arr] (d3) -- (d4);
  \draw[arr] (d4) -- (d5);
\end{tikzpicture}

%% file: sections/04_exact.tex
\section{Exact Coding: Replay, Lanes, and Routes}
\label{sec:exact}

\vex{} answers two questions per page: at which decimal scale may each value
enter the integer path, and how should the page's resolutions be grouped so
that the integers stay small?

\subsection{Replay-Checked Decimal Admission}
\label{sec:admission}

\begin{definition}[Admission]
\label{def:admission}
A value $x$ is \emph{admitted at scale $s$}, written $E_s(x)$, if
$x\otimes10^{s}$ is finite, $|q_s(x)|<2^{53}$, and
$\bits(R(q_s(x),s))=\bits(x)$.  Its \emph{minimal scale} $s^\ast(x)$ is the
least $s\le18$ with $E_s(x)$, and $q^\ast(x)=q_{s^\ast(x)}(x)$.
\end{definition}

The encoder evaluates Definition~\ref{def:admission} literally: for each value
and scale it computes $q_s(x)$, runs the decoder's division, and compares the
64-bit words, eight values per AVX-512 instruction, stopping at the first
match.  For $x=1.11$, $x\otimes10^{2}$ is $111.00000000000001$; the replay
rounds it to $q_2(x)=111$, finds $111\oslash10^{2}=x$, and admits $x$ at its
minimal scale~2 without asking whether the product is an integer.  An
admitted coordinate therefore reproduces its word by construction, and the
remaining question is how much the replay admits.

\begin{theorem}[Sufficient condition]
\label{thm:sufficient}
Let $0\le s\le18$ and $0<|N|<2^{51}$ an integer.  If $x=\fl(N\cdot10^{-s})$,
then $q_s(x)=N$ and $E_s(x)$ holds.
\end{theorem}

\begin{proof}
Both $N10^{-s}$ and $x10^{s}$ lie in the normal range, so
$x=N10^{-s}(1+\theta)$ and $x\otimes10^{s}=N(1+\theta)(1+\eta)$ with
$|\theta|,|\eta|\le2^{-53}$.  Hence $|x\otimes10^{s}-N|\le
(2^{51}-1)(2^{-52}+2^{-106})<\frac12$ and $q_s(x)=N$.  As $|N|<2^{53}$,
$\dbl(N)=N$ and $R(N,s)=\fl(N/10^{s})=x$; $x\ne0$ has the sign of $N$, so
the words agree.
\end{proof}

\begin{figure}[t]
\centering
\includegraphics[width=\columnwidth]{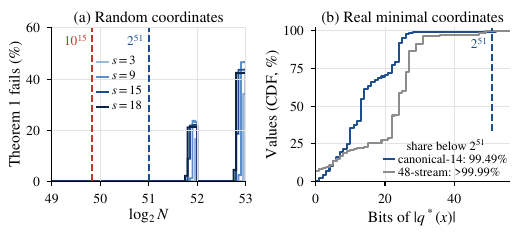}
\caption{Theorem~\ref{thm:sufficient} against measurement.  (a)~Share of
uniform coordinates $N$ for which $q_s(\fl(N\cdot10^{-s}))\ne N$ or the replay
fails, in bins of $2^{0.05}$.  (b)~Bit length of the minimal coordinate
$q^\ast(x)$ on the canonical pages and on the 48-stream suite.}
\label{fig:admission}
\end{figure}

Theorem~\ref{thm:sufficient} contains Falcon's guarantee for decimals with at
most 15 significant digits \cite{falcon} and reaches every 16-digit coordinate
below $2^{51}=2{,}251{,}799{,}813{,}685{,}248$, because it bounds both
roundings jointly against $\frac12$.  The bound is within a factor of two of
tight.  Over uniform coordinates (Figure~\ref{fig:admission}(a)), the
conclusion holds below $2^{51}$ at every scale and first fails between
$2^{51.79}$ and $2^{51.97}$; for instance, $x=\fl(4340765179169861\cdot10^{-15})$
has $q_{15}(x)=N+1$ and fails the replay at $s=15$.  The replay itself needs no
bound: it admits the same $x$ at $s=16$, and it rejects exactly the words that
would not survive.  For $x=-0.0$, $R(0,s)=+0.0$, so $-0.0$ is never admitted,
and NaN and $\pm\infty$ fail finiteness.  On real data the theorem covers
almost everything the replay admits (Figure~\ref{fig:admission}(b)): the
median minimal coordinate has 13 bits on the canonical pages and 24 bits on
the 48 streams, and 99.49\% and all admitted values lie below $2^{51}$.  The
other 0.51\% are Air-sensor coordinates between $2^{51}$ and $2^{53}$, 86.5\%
of that stream's admitted values, which the replay admits value by value.
The replay admits all but 1,264 of the 1,184,768 canonical values (1,231 long
fractions of Air-sensor and 33 negative zeros) and all but 120 of the 74.7
million values of the 48 streams; these take the binary lane or stay as raw
words.

\subsection{Lanes, Lifting, and Routes}
\label{sec:lanes}

A one-scale codec places every value of a page at the page's finest scale.
The next theorem says this is always correct and Proposition~\ref{prop:penalty}
says what it costs.

\begin{theorem}[Lifting]
\label{thm:lifting}
If $E_s(x)$ holds with $q=q_s(x)$ and $|10^{k}q|\le2^{53}$ for some $k\ge0$
with $s+k\le18$, then $R(10^{k}q,\,s+k)=x$.
\end{theorem}

\begin{proof}
$10^{k}q$, $q$, $10^{s}$, and $10^{s+k}$ are binary64 values and the exact
quotients $10^{k}q/10^{s+k}$ and $q/10^{s}$ are equal, so their correctly
rounded quotients are equal.
\end{proof}

\begin{definition}[Lanes and route]
\label{def:lanes}
A \emph{lane plan} is a list of scales $S_1<\cdots<S_L$ with $L\le4$.  Lane
$j$ holds, in page order, every admitted value with
$S_{j-1}<s^\ast(x)\le S_j$ ($S_0=-1$) at coordinate
$u=10^{S_j-s^\ast(x)}q^\ast(x)$ when $|u|\le2^{53}$.  The \emph{route} lists
the lane of every position.  Values that no lane holds are
\emph{exceptions}.
\end{definition}

\textbf{Example.}  The page of Figure~\ref{fig:strategies} holds four values of
minimal scale~3 and two of scale~5.  The single lane $\langle5\rangle$ codes
$1257900,1258900,1259400,1259621,1259900,1260381$, whose steps $+1000,+500$
carry the lift.  The plan $\langle3,5\rangle$ routes
\texttt{C\,C\,C\,F\,C\,F}: lane~1 holds $12579,12589,12594,12599$, whose steps
$+10,+5,+5$ share the gcd~5 and are coded as $+2,+1,+1$, and lane~2 holds
$1259621,1260381$ with the single step $+760$.

By Theorem~\ref{thm:lifting}, the decoder restores every lane value with the
same division $R(u,S_j)$, so lane plans differ only in size.  Lifting is
expensive because it multiplies steps.  The following lemma is immediate from
$\zz(Md)\ge M\zz(d)$ for $M\ge1$.

\begin{lemma}[Length shift]
\label{lem:shift}
For integers $d\ne0$ and $M\ge1$,
$\ell(\zz(Md))\ge\ell(\zz(d))+\lfloor\log_2M\rfloor$.
\end{lemma}

\begin{proposition}[Lifting penalty]
\label{prop:penalty}
Let a lane at scale $S$ hold two consecutive values of minimal scale $s<S$
whose scale-$s$ step is $d'\neq0$, and let the lane's step gcd $g$ be coprime
to $10$.  Then the coded step is $10^{S-s}d'/g$, at least
$\lfloor(S-s)\log_2 10\rfloor$ bits longer than $d'/g$.
\end{proposition}

\begin{proof}
By Definition~\ref{def:lanes} the step is $10^{S-s}d'$; $g$ divides it and is
coprime to $10^{S-s}$, so $g\mid d'$.  Apply Lemma~\ref{lem:shift} with
$M=10^{S-s}$.
\end{proof}

The gcd is coprime to $10$ whenever the fine values make the last digit
arbitrary, which is the case that forces the lift.  A 0.1-unit reading lifted
to a 0.001-unit lane thus pays $\lfloor2\log_210\rfloor=6$ bits per nonzero
step, and in an 8-value block a single lifted step widens all eight slots.
A lane of its own removes this penalty, and since each lane differences its
own history, the fine values no longer interrupt the coarse sequence.  The
price is the route, a stream of lane indices: rare fine values become sparse
exceptions of a constant route, and a burst fills whole 8-value blocks with
one index.  Two kinds of value need more than a decimal lane.

\textbf{Rational lattices.}  Averages and unit conversions land on grids such
as thirds or sevenths of a decimal unit; Appliances-RH stores values such as
$36.2128571428571\ldots=25349/700$.  A decimal lane at scale $S$ may
therefore switch to a rational lattice with $\nu=10^{S}$ and
$\mu=k\cdot10^{c}$, where $k\in\{3,6,7,9,11,12,13,14,15,18,21,24,36\}$,
$c\le4$, and $\nu/\mu\ge8$.  It codes $\kappa_i=\rint(u_i\mu/\nu)$ and
residues $r_i=u_i-\varphi(\kappa_i)$, where
$\varphi(\kappa)=\rint(\kappa\nu/\mu)$ is evaluated in exact integer arithmetic
by splitting $\kappa=a\mu+b$.  The decoder computes $\varphi(\kappa_i)+r_i=u_i$
and divides as before, so every grid decodes exactly, and on-grid values have
$r_i=0$ and a constant residue stream.  Since
$|\Delta u|\ge|\Delta\kappa|\,\nu/\mu-1$, Lemma~\ref{lem:shift} shows that
coding $\kappa$-steps saves about $\log_2(\nu/\mu)$ bits per nonzero step.  The
encoder samples 64 coordinates, screens each candidate grid by fractional
parts, and replays $\varphi$ on the survivors.

\textbf{Binary lane and raw words.}  Exceptions are words that no decimal lane
holds: signed zeros, NaN, $\pm\infty$, long binary fractions, and lifts past
$2^{53}$.  With at least eight of them, a binary lane uses the 64-bit words as
coordinates, with differences modulo $2^{64}$, so float32-origin words share
a power-of-two step gcd that the stream coder factors out.  Otherwise each
exception keeps its 64-bit word, with gamma-coded gaps or a bitmap for its
positions.

\subsection{Planning: \fast{} and \upper{}}
\label{sec:planning}

A plan is a scale list plus, per lane, a \emph{variant} (first or second
differences, decimal or rational).  Both profiles start from the single lane
at the page's largest minimal scale, which Theorem~\ref{thm:lifting} makes
feasible, and insert one scale at a time (Algorithm~\ref{alg:plan}).

\begin{algorithm}[t]
\caption{\vex{} page encoder}
\label{alg:plan}
\footnotesize
\begin{algorithmic}[1]
\Require page $x_0,\ldots,x_{n-1}$; profile $\in\{\fast,\upper\}$
\State $(s^\ast_i,q^\ast_i)\gets$ least $s\le18$ with $E_s(x_i)$ \Comment{vector replay}
\State $\Sigma\gets\{s^\ast_i\}$; $P\gets\langle\max\Sigma\rangle$; $c\gets\textsc{Price}(P)$
\If{profile $=\upper$}
  \While{$|P|<4$}
    \State $(P',c')\gets\arg\min_{s\in\Sigma\cup\{\mathrm{bin}\}\setminus P}
      \textsc{Price}(P\cup\{s\})$ \Comment{cached per interval}
    \If{$c'\ge c$} \textbf{break} \EndIf
    \State $(P,c)\gets(P',c')$
  \EndWhile
\Else
  \State $P'\gets$ greedy over $\Sigma$ by the block-width proxy, $\le3$ lanes
  \If{$\neg$(lane of $P$ rational and complete) $\wedge\textsc{Price}(P')<c$}
    \State $(P,c)\gets(P',\textsc{Price}(P'))$
  \EndIf
  \If{$\textsc{Price}(P\cup\{\mathrm{bin}\})<c$} $P\gets P\cup\{\mathrm{bin}\}$ \EndIf
\EndIf
\State emit $P$; if not shorter than $1+8n$ bytes, emit the raw page
\Statex \textsc{Price}$(P)$: gather lanes, route, and exceptions; per lane,
  the cheapest variant under the stream coder; sum all bits.
\end{algorithmic}
\end{algorithm}

\upper{} prices every candidate exactly: it gathers the lanes, prices the
header, route, and exceptions, and lets the stream coder of
Section~\ref{sec:coder} price each lane under each variant.  It inserts the
present scale, or the binary lane, that lowers the page size most, and stops
when no insertion helps or four lanes are reached.  Lane prices depend only on
the scale interval, so a cache holds at most $19\cdot20/2$ intervals per page.

\fast{} ranks plans with a proxy.  A lane costs eight times the zigzag width of
the widest first difference in each 8-value block, less the gcd's bits, the
route costs its empirical entropy, and each value left out costs 72 bits.  The
proxy inserts up to three decimal lanes, each only when it improves the proxy
by 0.5\%.  \fast{} then prices the proxy's plan exactly against the single lane
and keeps the cheaper one, skipping the multi-lane plan when the single lane is
rational and holds every value, and finally adds the binary lane if that
lowers the exact price.

\begin{proposition}[Never worse than one lane]
\label{prop:monotone}
Under both profiles, the planned size of a page is at most that of the single
lane at its largest minimal scale, and every emitted page has at most $1+8n$
bytes.
\end{proposition}

\begin{proof}
Both planners begin with the single lane priced exactly and replace it only
by a plan with a lower exact price; the emitter falls back to a raw page,
one tag byte and the $n$ words, whenever the lane page is not shorter.
\end{proof}

\upper{} evaluates at most 80 candidate plans per page, each in $O(n)$, and
\fast{} prices at most three plans exactly; both encoders are therefore linear
in $n$, and the constant separates their speeds
(Section~\ref{sec:evaluation}).

%% file: sections/05_bounded.tex
\section{Bounded Coding: FMA Replay on an Error Lattice}
\label{sec:bounded}

Under the $\varepsilon$-bounded contract a value may move by less than
$\varepsilon$, so \veb{} chooses a lattice $o+\delta\zz$ per page, codes the
index of each value's lattice point, and checks the bound with the decoder's
own arithmetic.

\textbf{Lattice and replay.}  The step is the binary64 value
$\delta=2\varepsilon(1-2^{-40})$, so a lattice point lies within
$\delta/2<\varepsilon$ of every real value, with a margin of
$\varepsilon2^{-40}$ for rounding.  For a value $x$ the encoder computes
$t=\fl(\fl(x-o)\cdot\fl(1/\delta))$, $q=\rint(t)$, and the decoder's
reconstruction $\hat x=\mathrm{fma}(q,\delta,o)$.  It admits $x$ if
\begin{equation}
\label{eq:accept}
\fl(|x-\hat x|)\le\mathrm{pred}(\varepsilon)\quad\text{and}\quad|q|<2^{50},
\end{equation}
where $\mathrm{pred}(\varepsilon)$ is the binary64 predecessor of
$\varepsilon$.  When the first index fails, it tries $q\pm1$ and $q\pm2$, which
recovers values whose quotient $t$ was rounded across a cell boundary.  A value
that still fails, and every non-finite value, is an exception and keeps its
64-bit word.

\begin{theorem}[Bounded replay]
\label{thm:bounded}
Every admitted value satisfies $|\hat x-x|<\varepsilon$, and \veb{} is
$\varepsilon$-bounded in the sense of Definition~\ref{def:contracts}.
\end{theorem}

\begin{proof}
$\fl$ is monotone and $\varepsilon$ is a binary64 value, so
$|x-\hat x|\ge\varepsilon$ would give $\fl(|x-\hat x|)\ge\varepsilon>
\mathrm{pred}(\varepsilon)$, violating Equation~\eqref{eq:accept}.  The
decoder evaluates the same correctly rounded $\mathrm{fma}(q,\delta,o)$ with
the same binary64 operands, so it outputs the checked $\hat x$; exceptions
are restored word for word.
\end{proof}

The proof needs no error analysis of $t$ or of the quotient, because the
replay checks the decoder's own result and so no decoder rounding has to be
budgeted.  Error analysis instead bounds how rarely the replay rejects a
value.

\begin{proposition}[Admission margin]
\label{prop:margin}
Let $x$ be finite with $|x-o|<2^{49}\delta$, and let $p=q^\ast\delta+o$ be the
lattice point nearest to $x$.  The encoder tries $q^\ast$, and admits $x$
whenever $|x-p|\le\mathrm{pred}(\varepsilon)-\frac12\mathrm{ulp}(p)$.  Hence $x$
can fail only within $\frac12\mathrm{ulp}(p)$ of a cell boundary, a share of
at most $\mathrm{ulp}(p)/\delta$ of each cell, and every $x$ with
$|p|\le2^{12}\varepsilon$ is admitted.
\end{proposition}

\begin{proof}
$t$ is $r=(x-o)/\delta$ after three roundings, so $|t-r|\le3.01\cdot2^{-53}|r|<\frac14$
and $\rint(t)$ is $q^\ast$ or a neighbor, all of which the encoder tries.  The
fused multiply-add returns $\fl(p)$ with $|\fl(p)-p|\le\frac12\mathrm{ulp}(p)$,
so $|x-\hat x|\le\mathrm{pred}(\varepsilon)$ under the stated condition, and
monotone rounding keeps $\fl(|x-\hat x|)\le\mathrm{pred}(\varepsilon)$.  Since
$|x-p|\le\delta/2\le\varepsilon(1-2^{-40}+2^{-53})$ and
$\mathrm{pred}(\varepsilon)\ge\varepsilon(1-2^{-52})$, the condition holds
whenever $\frac12\mathrm{ulp}(p)\le2^{-41}\varepsilon$, which
$|p|\le2^{12}\varepsilon$ implies.
\end{proof}

For typical magnitudes the rejection band is negligible: at
$\varepsilon=10^{-6}$ and $|x|\approx10^{3}$ it covers about $10^{-7}$ of each
cell.  Over all 90 cells, 10.5 million values, no value became an exception.

\begin{figure}[t]
\centering
\resizebox{\columnwidth}{!}{\input{figures/fig_origins}}
\caption{Origins of the bounded lattice.  Solid ticks are lattice points,
dashed lines cell boundaries, and dots one page's values; $q$ lists their
indices in page order.}
\label{fig:origins}
\end{figure}

\textbf{Phase and origin.}  A lattice through zero may place a page's values
near cell boundaries, where a small drift flips $q$ and widens the steps
(Figure~\ref{fig:origins}(a)).
\veb{} therefore considers three origins.  The zero origin costs nothing.  A
phase origin $o=\varphi\,\delta/256$ with an 8-bit $\varphi$ is taken from the
first finite value's offset in its cell when at least a quarter of the page
clusters within $\delta/32$ of that offset, which moves a lattice point onto
the cluster (Figure~\ref{fig:origins}(b)); the encoder prices both lattices
and keeps the cheaper.  A page whose range fits in $0.97\delta$ is centered on
its midrange, so every finite value maps to one point and the index stream is
constant (Figure~\ref{fig:origins}(c)).  When $\max|x|/\delta\ge2^{49}$, the origin is the first finite
value itself, written as 64 bits, which keeps every index below $2^{50}$ for
pages that span far less than their magnitude, such as coordinates and prices.

\textbf{Indices to steps.}  The admitted indices, in page order, form one
integer sequence whose first differences go to the stream coder.  Exception
positions are gamma-coded gaps or a bitmap, whichever is shorter, and a page
that would exceed $1+8n$ bytes is written raw.  Because the index sequence of
a bounded page and a lane of an exact page are both integer sequences with
steps, the stream coder of the next section serves both without change, and
the same factoring that finds a decimal lane's gcd finds a bounded page's
coarse quantization.

\textbf{Example.}  A thermometer with resolution $0.1$ reports $21.4$,
$21.5$, $21.5$, $21.7$, bounded at $\varepsilon=0.01$.  With the zero origin
and $\delta\approx0.02$ the indices are $1070,1075,1075,1085$, each replayed
within $10^{-10}$ of its value, and the steps $+5,0,+10$ share the gcd~5.
The stream coder stores $+1,0,+2$, zigzag $2,0,4$, in 3-bit blocks, the same
bits as the exact lane at scale~1 for this page: the lattice is finer than
the data, and the gcd recovers the data's own resolution.

%% file: figures/fig_origins.tex
% Origins. Zoomed cells, Falcon Fig. 7 number-line grammar.
\begin{tikzpicture}[x=1cm,y=1cm,
  every node/.style={inner sep=0pt, outer sep=0pt},
  title/.style={font=\scriptsize\bfseries, anchor=west},
  ticklab/.style={font=\fontsize{6.4}{7.2}\selectfont, anchor=north, inner sep=1pt},
  colh/.style={font=\fontsize{6.4}{7.2}\selectfont, text=black!55, anchor=center},
  qv/.style={font=\scriptsize, anchor=center},
  axline/.style={draw=black!72, line width=0.55pt},
  lat/.style={draw=cC, line width=0.95pt},
  bnd/.style={draw=black!45, dash pattern=on 1.1pt off 0.9pt, line width=0.55pt}]
\def\xA{0.08}\def\xB{5.18}
\def\tL{0.36}\def\tR{4.90}\def\tM{2.63}
\def\qX{6.22}\def\sX{7.78}
% column heads
\node[colh] at (\qX,0.02) {$q$};
\node[colh] at (\sX,0.02) {steps};
% ---------- (a) ----------
\node[title] at (0.02,0) {(a) Zero origin};
\draw[axline] (\xA,-0.34) -- (\xB,-0.34);
\draw[lat] (\tL,-0.44) -- (\tL,-0.20);
\draw[lat] (\tR,-0.44) -- (\tR,-0.20);
\draw[cF, dash pattern=on 1.1pt off 0.9pt, line width=0.7pt] (\tM,-0.50) -- (\tM,-0.16);
\foreach \x in {2.05,2.40,2.86,3.21} {\fill (\x,-0.34) circle (1.28pt);}
\node[ticklab] at (\tL,-0.46) {$2\delta$};
\node[ticklab] at (\tR,-0.46) {$3\delta$};
\node[qv] at (\qX,-0.36) {$2,3,2,3$};
\node[qv, text=cF] at (\sX,-0.36) {$+1,-1,+1$};
% ---------- (b) ----------
\node[title] at (0.02,-0.88) {(b) Phase origin $o=\varphi\delta/256$};
\draw[axline] (\xA,-1.20) -- (\xB,-1.20);
\draw[bnd] (\tL,-1.34) -- (\tL,-1.06);
\draw[bnd] (\tR,-1.34) -- (\tR,-1.06);
\draw[lat] (\tM,-1.38) -- (\tM,-1.00);
\foreach \x in {2.05,2.40,2.86,3.21} {\fill (\x,-1.20) circle (1.28pt);}
\node[ticklab] at (\tM,-1.36) {$o+2\delta$};
\node[qv] at (\qX,-1.22) {$2,2,2,2$};
\node[qv, text=cC!75!black] at (\sX,-1.22) {$0,0,0$};
% ---------- (c) ----------
\node[title] at (0.02,-1.76) {(c) One-cell page, $o$ = midrange};
\draw[axline] (\xA,-2.08) -- (\xB,-2.08);
\draw[bnd] (\tL,-2.22) -- (\tL,-1.94);
\draw[bnd] (\tR,-2.22) -- (\tR,-1.94);
\draw[lat] (\tM,-2.26) -- (\tM,-1.88);
\foreach \x in {0.78,1.62,3.58,4.48} {\fill (\x,-2.08) circle (1.28pt);}
\draw[black!60, {Stealth[length=2pt]}-{Stealth[length=2pt]}] (0.78,-2.36) -- (4.48,-2.36);
\node[ticklab, anchor=north] at (\tM,-2.38) {range $\le 0.97\delta$};
\node[qv] at (\qX,-2.10) {$0,0,0,0$};
\node[qv, text=cC!75!black] at (\sX,-2.10) {constant};
\path[use as bounding box] (0,0.14) rectangle (8.72,-2.72);
\end{tikzpicture}

%% file: sections/06_coder.tex
\section{The Shared Integer Stream Coder}
\label{sec:coder}

\begin{figure}[t]
\centering
\resizebox{\columnwidth}{!}{\input{figures/fig_format}}
\caption{Page and stream layouts.  An exact lane, the route, rational and
split residues, and a bounded page's index steps are all streams in the
bottom layout; the header of every stream names its representation.}
\label{fig:format}
\end{figure}

Both front ends end in integer sequences: lane coordinates, rational indices
and residues, and the route under the exact contract; lattice indices under
the bounded one.  A sequence $u_0,\ldots,u_{m-1}$ is written as its anchor
$u_0$ and the steps $z_i=u_{i+1}-u_i$ (or second differences, in \upper{}),
and the step stream passes through three stages.

\textbf{Exact gcd.}  The coder divides all steps by $g=\gcd_i|z_i|$, which
removes the lifting factor of a lane whose values all share a coarser scale,
the power of two of a float32-origin binary lane, and the coarse quantization
of a bounded page; lane~1 of Figure~\ref{fig:strategies} and the thermometer
of Section~\ref{sec:bounded} both have $g=5$.  The gcd is maintained
incrementally: each step is tested
for divisibility by $g$ through the floating-point reciprocal and an exact
integer multiply-back, eight steps per instruction, and a step that fails
updates $g$ by a binary gcd.  The test never accepts a false divisor because
the multiply-back is exact, so the division of the accepted steps is exact.

\textbf{Coarse/residue split.}  Some sequences are a coarse lattice plus a
small jitter, such as readings that drift around a calibration grid or
trajectories at fine bounds.  From the three most frequent large steps of a
256-step sample the coder derives candidate lattices $g_2$; if one places at
least 80\% of the sampled steps within $g_2/64$ of its multiples, the
coordinate $s_i=\sum_{j\le i}z_j$ is split exactly as $s_i=g_2k_i+\rho_i$, and
the coarse steps $k_{i}-k_{i-1}$ and the residues $\rho_i$ become two streams.
The split is kept only when its exact price is lower.

\textbf{Four representations.}  Each stream is priced, in bits, under four
page-local representations and written in the cheapest:
\begin{itemize}[leftmargin=*,nosep]
\item \emph{Base}: a constant, or a base value with at most $m/16$
exceptions, coded as Exp-Golomb gaps and offsets.  Constant residues and
regular steps cost a few bits per page.
\item \emph{Blocks}: frame-of-reference over 8-value blocks of zigzag steps,
each block at the width of its widest step (at most 56 bits), with
Exp-Golomb-coded width changes.
\item \emph{Hybrid rANS}: four interleaved rANS states \cite{rans} over 138
symbols: 16 literal zigzag values, and escape classes indexed by bit length
and the next bit, followed by the remaining bits verbatim.
\item \emph{Dictionary rANS}: when at most 256 distinct steps occur, rANS over
their indices, with the sorted dictionary coded by gaps.
\end{itemize}
With $v_i=\zz(z_i)$, $\ell$ the bit length, and $f$ a frequency table
normalized deterministically to $2^{p}$ ($p\le11$, chosen from $m$), the
prices are
\begin{align*}
C_{\mathrm{blk}}&=\textstyle\sum_{b}8\,\ell\big(\max_{i\in b}v_i\big)+C_{\mathrm{EG}}(\Delta w)+c_1,\\
C_{\mathrm{hyb}}&=\textstyle\sum_{i}\big(p-\log_2f_{\sigma(v_i)}+e(v_i)\big)+C_{f}+c_2,\\
C_{\mathrm{dic}}&=\textstyle\sum_{i}\big(p-\log_2f_{\iota(z_i)}\big)+C_{\mathrm{EG}}(\text{gaps})+C_{f}+c_2,\\
C_{\mathrm{base}}&=\textstyle C_{\mathrm{EG}}(\text{gaps})+C_{\mathrm{EG}}(\text{offsets})+\gamma(k+1)+c_3,
\end{align*}
where $\sigma$ maps a value to its hybrid symbol, $e(v)=\ell(v)-2$ for
$v\ge16$ and $0$ otherwise counts its escape bits, $\iota$ maps a step to its
dictionary index, $C_f$ is the coded table, and the constants cover headers
and the four 32-bit rANS states.  The decoder uses the same normalized tables, so prices
match the written size up to rANS renormalization and byte padding; over the
canonical pages the planned bits are within 0.3\% of the emitted bytes.
\fast{} disables both rANS representations, which removes entropy decoding
from its pages; \veb{} and \upper{} use all four.

\textbf{Decoding.}  The stream header names the representation, the gcd, and
the split, so the decoder performs no search.  Blocks unpack with vector shifts
and masks; rANS decodes four states in parallel and resolves hybrid escapes
eight at a time with a byte permute and a funnel shift; a vector prefix sum
then rebuilds $u_{i+1}=u_0+g\,s_i$, with $s_i=g_2k_i+\rho_i$ under a split,
and a lane converts its
coordinates to doubles and divides by $10^{S}$ in the same pass, or evaluates
$\mathrm{fma}(q,\delta,o)$ for a bounded page.  The route scatters lane values
back to page positions, and exceptions overwrite their positions last.

\begin{theorem}[Page round trip]
\label{thm:page}
For every page $X$, \vex{} restores $\bits(x_i)$ for all $i$, and \veb{}
restores every finite $x_i$ within $\varepsilon$ and every other word exactly.
\end{theorem}

\begin{proof}
Each stage is an exact integer bijection given its header: zigzag and the
representations are invertible, the gcd divides every step exactly, the split
is an integer identity, and prefix sums invert differences modulo $2^{64}$.
Hence the decoder recovers every coordinate, rational index and residue,
route, and lattice index.  For \vex{}, Theorem~\ref{thm:lifting} and the
admission check give $\bits(R(u,S))=\bits(x)$ for decimal lanes, $\varphi$ is
exact integer arithmetic, binary-lane coordinates are the words, and
exceptions are stored verbatim.  For \veb{}, Theorem~\ref{thm:bounded}
applies.  A raw page stores the words.
\end{proof}

\textbf{Cost.}  Pricing is linear in $m$: one pass for the gcd, one for the
block widths, one histogram for the hybrid model, and a hash table for the
dictionary that gives up once half of the first 128 steps are distinct.
Decoding is linear with no data-dependent search, and pages are independent,
so both directions parallelize across pages.

%% file: figures/fig_format.tex
% Page and stream layouts. Ruled bars, Falcon Fig. 2 / Elf Fig. 7.
\begin{tikzpicture}[x=1cm,y=1cm,
  every node/.style={inner sep=0.5pt, outer sep=0pt, font=\fontsize{6.5}{7.4}\selectfont},
  lab/.style={font=\scriptsize\bfseries, anchor=east, inner sep=1pt},
  rule/.style={draw=black!58, line width=0.4pt}]
\def\xL{1.38}\def\xR{8.68}
\def\bh{0.46}
% breaks, exact / bounded share the first three fields
\def\a{1.96}   % after tag
\def\b{4.22}   % after meta / origin
\def\c{5.48}   % after exceptions
\def\d{6.26}   % after route
\def\e{7.12}   % after lane 1
\def\f{7.50}   % after cdots
\def\g{6.22}   % after anchor (bounded)
% ---------- exact ----------
\fill[cC!6] (\xL,0) rectangle (\xR,-\bh);
\fill[cC!16] (\c,0) rectangle (\xR,-\bh);
\draw[rule] (\xL,0) rectangle (\xR,-\bh);
\foreach \x in {\a,\b,\c,\d,\e,\f} {\draw[black!32] (\x,0) -- (\x,-\bh);}
\node at ({(\xL+\a)/2},-0.23) {tag};
\node at ({(\a+\b)/2},-0.23) {$L$, scales, variants};
\node at ({(\b+\c)/2},-0.23) {exceptions};
\node at ({(\c+\d)/2},-0.23) {route};
\node at ({(\d+\e)/2},-0.23) {lane 1};
\node at ({(\e+\f)/2},-0.23) {$\cdots$};
\node at ({(\f+\xR)/2},-0.23) {lane $L$};
\node[lab] at (\xL-0.08,-0.23) {Exact};
% ---------- bounded ----------
\def\yB{-0.52}
\fill[cF!6] (\xL,\yB) rectangle (\xR,\yB-\bh);
\fill[cF!15] (\g,\yB) rectangle (\xR,\yB-\bh);
\draw[rule] (\xL,\yB) rectangle (\xR,\yB-\bh);
\foreach \x in {\a,\b,\c,\g} {\draw[black!32] (\x,\yB) -- (\x,\yB-\bh);}
\node at ({(\xL+\a)/2},\yB-0.23) {tag};
\node at ({(\a+\b)/2},\yB-0.23) {origin: $0$ / $\varphi$ / $o$};
\node at ({(\b+\c)/2},\yB-0.23) {exceptions};
\node at ({(\c+\g)/2},\yB-0.23) {anchor};
\node at ({(\g+\xR)/2},\yB-0.23) {index steps};
\node[lab] at (\xL-0.08,\yB-0.23) {Bounded};
% ---------- stream ----------
\def\yS{-1.06}
\fill[black!3] (\xL,\yS) rectangle (\xR,\yS-\bh);
\def\sA{2.28}\def\sB{2.70}\def\sC{3.82}\def\sD{5.36}\def\sE{7.28}
\fill[black!7] (\sD,\yS) rectangle (\sE,\yS-\bh);
\draw[rule] (\xL,\yS) rectangle (\xR,\yS-\bh);
\foreach \x in {\sA,\sB,\sC,\sD,\sE} {\draw[black!32] (\x,\yS) -- (\x,\yS-\bh);}
\node at ({(\xL+\sA)/2},\yS-0.23) {anchor};
\node at ({(\sA+\sB)/2},\yS-0.23) {$g$};
\node at ({(\sB+\sC)/2},\yS-0.23) {split $g_2$};
\node at ({(\sC+\sD)/2},\yS-0.23) {mode, params};
\node at ({(\sD+\sE)/2},\yS-0.23) {body};
\node at ({(\sE+\xR)/2},\yS-0.23) {extra bits};
\node[lab] at (\xL-0.08,\yS-0.23) {Stream};
% ---------- mode ----------
\def\yM{-1.60}
\def\mh{0.52}
\fill[black!2] (\xL,\yM) rectangle (\xR,\yM-\mh);
\draw[rule] (\xL,\yM) rectangle (\xR,\yM-\mh);
\def\mA{3.205}\def\mB{5.03}\def\mC{6.855}
\foreach \x in {\mA,\mB,\mC} {\draw[black!32] (\x,\yM) -- (\x,\yM-\mh);}
\node[align=center] at ({(\xL+\mA)/2},\yM-0.26) {base\\[-1pt]{\fontsize{5.6}{6.4}\selectfont constant / sparse}};
\node at ({(\mA+\mB)/2},\yM-0.26) {8-value blocks};
\node at ({(\mB+\mC)/2},\yM-0.26) {hybrid rANS};
\node at ({(\mC+\xR)/2},\yM-0.26) {dictionary rANS};
\node[lab] at (\xL-0.08,\yM-0.26) {Mode};
\path[use as bounding box] (0,0.12) rectangle (8.72,-2.16);
\end{tikzpicture}

%% file: sections/07_evaluation.tex
\section{Experimental Evaluation}
\label{sec:evaluation}

We ask five questions.  \textbf{Q1:} how small and fast are exact pages
against exact baselines on the same page boundary?  \textbf{Q2:} what changes
inside a storage engine?  \textbf{Q3:} how do mixed resolutions affect each
codec?  \textbf{Q4:} how does \veb{} compare with error-bounded codecs?
\textbf{Q5:} which mechanisms carry the results?

\subsection{Setup}
\label{sec:eval-setup}

\textbf{Exact workloads.}  The canonical corpus is the 14-stream Elf/Elf*
suite, byte-identical to the SElfStar artifact \cite{elfstar}.  The 48-stream
suite holds 74,697,728 values (598~MB) from UCI household power (7 streams)
\cite{uci_household}, Beijing air quality (11) \cite{beijingair}, appliance
energy sensors (26) \cite{appliances}, and GeoLife and T-Drive trajectories
(4) \cite{geolife,tdrive}.  Every codec resets at each complete 1,024-value
page.  The storage workload writes four of these streams (6,622,208 values)
into Apache TsFile \cite{tsfile}, and the compaction workload writes all 48.

\textbf{Exact baselines.}  ALP, ELF*, ELF, Chimp, Gorilla, and DeXOR run
from the authors' C++ code \cite{alp,elfstar,elf,chimp,gorilla,dexor} in one
harness; Falcon runs on an NVIDIA RTX PRO 6000 GPU with CUDA~12.8, and its
rates include host--device copies \cite{falcon}; Pcodec~1.0.4 \cite{pcodec}
runs through its Python binding, so we report its ratio only, as for xz~5.2.5
and zstd~1.5.7 \cite{zstd}, which compress each raw page on its own.

\textbf{Bounded workload.}  We use the 15 streams of the Serf benchmark with
absolute bounds $\varepsilon\in\{10^{-1},\ldots,10^{-6}\}$, 90 cells in
total, and 1,024-value pages for every codec.  Machete runs from the Serf
repository \cite{serf,machete} with its Lorenzo predictor and hybrid coder in
the same process as \veb{}; Serf-QT, SZ2, SZ3~3.2.1, ZFP~1.0.1, Sim-Piece,
and Sprintz \cite{serf,sz,zfp,simpiece,sprintz} come from the same harness on
the same pages.

\textbf{Implementation.}  \ravel{} is about 4,100 lines of C++17 over the
standard library, and the TsFile integration is a thin shim around the page
functions.

\textbf{Protocol.}  Single-thread runs are pinned to one core of a 64-core AMD
EPYC 9554 (Zen~4, 256~GB memory, Ubuntu~22.04, Linux~6.8) with GCC~12.3,
\texttt{-O3 -march=native -ffp-contract=off}.  Each measurement is the median
of five (canonical), three (48 streams), or eleven (bounded) timed runs after a
verified warm-up, with data in memory.  Over ten repeated canonical runs,
\ravel{}'s aggregate encode rates vary by at most 5\% and its decode rates by
2\%, and its output bytes never change.  Ratio is raw bytes over all output
bytes; throughput is raw MB/s.  Every \ravel{} output is checked word by
word, or against $\varepsilon$, and none of its numbers uses a secondary
compressor.

\begin{table*}[t]
\centering
\caption{Canonical streams with independent 1,024-value pages: compression
ratio and one-thread throughput (higher is better).  Bold marks the best
ratio and the fastest rate.  Pcodec is ratio only.}
\label{tab:canonical}
\footnotesize
\setlength{\tabcolsep}{2.0pt}
\renewcommand{\arraystretch}{0.92}
\begin{tabular}{@{}l*{15}{r}@{}}
\toprule
 & CT & IR & WS & P10 & SUK & USA & SDE & DT & AP & BW & BT & BP & BM & AS & All\\
\midrule
\multicolumn{16}{@{}l}{\itshape Compression ratio}\\
\vex{} \upper{} & \textbf{9.66} & \textbf{17.44} & \textbf{11.99} & \textbf{18.93} & \textbf{12.93} & \textbf{13.82} & \textbf{12.79} & \textbf{7.59} & \textbf{7.28} & \textbf{2.57} & \textbf{5.61} & \textbf{3.01} & \textbf{5.95} & 1.37 & \textbf{7.53}\\
\vex{} \fast{} & 8.92 & 16.98 & 11.89 & 17.53 & 12.42 & 13.36 & 12.29 & 7.56 & 6.42 & 2.40 & 4.54 & 3.00 & 5.76 & 1.34 & 7.01\\
Falcon & 8.18 & 15.90 & 11.36 & 13.82 & 8.99 & 12.25 & 8.86 & 7.21 & 6.09 & 2.09 & 2.16 & 2.91 & 4.03 & 1.34 & 5.45\\
Pcodec & 7.95 & 12.61 & 10.20 & 12.96 & 10.13 & 10.81 & 10.02 & 6.35 & 5.17 & 1.95 & 1.65 & 1.37 & 3.63 & \textbf{1.38} & 4.82\\
ELF* & 5.88 & 7.69 & 6.10 & 9.42 & 6.52 & 6.93 & 5.22 & 4.79 & 4.31 & 1.99 & 2.15 & 2.36 & 2.97 & 1.27 & 4.20\\
ELF & 4.80 & 6.43 & 5.09 & 8.43 & 5.44 & 5.70 & 4.53 & 4.07 & 4.00 & 1.83 & 1.94 & 2.10 & 2.72 & 1.18 & 3.70\\
ALP & 6.00 & 6.74 & 9.45 & 7.02 & 5.91 & 7.06 & 5.53 & 5.11 & 3.79 & 2.03 & 2.15 & 2.44 & 2.97 & 1.25 & 4.19\\
Gorilla & 1.17 & 1.43 & 1.21 & 2.08 & 1.75 & 1.48 & 1.39 & 1.20 & 1.40 & 1.01 & 1.06 & 1.20 & 1.27 & 1.22 & 1.32\\
Chimp & 1.56 & 1.56 & 1.22 & 2.34 & 1.93 & 1.56 & 1.49 & 1.29 & 1.53 & 1.14 & 1.18 & 1.31 & 1.40 & 1.29 & 1.46\\
DeXOR & 5.64 & 7.41 & 5.92 & 9.15 & 5.51 & 6.57 & 5.22 & 4.83 & 4.31 & 2.09 & 2.19 & 2.48 & 3.28 & 1.22 & 4.20\\
\midrule
\multicolumn{16}{@{}l}{\itshape Encode MB/s (Falcon on CUDA)}\\
\vex{} \upper{} & 151 & 150 & 159 & 123 & 130 & 128 & 97 & 129 & 129 & 25 & 41 & 124 & 49 & 66 & 78\\
\vex{} \fast{} & 625 & \textbf{849} & \textbf{908} & \textbf{708} & \textbf{818} & \textbf{831} & \textbf{712} & \textbf{691} & 591 & 197 & 175 & 607 & 497 & 299 & 476\\
Falcon & 48 & 49 & 55 & 46 & 46 & 45 & 43 & 45 & 42 & 29 & 27 & 14 & 36 & 11 & 40\\
ELF* & 159 & 147 & 150 & 155 & 137 & 148 & 140 & 151 & 158 & 115 & 122 & 141 & 114 & 149 & 140\\
ELF & 180 & 168 & 163 & 186 & 151 & 164 & 157 & 171 & 193 & 148 & 161 & 171 & 140 & 161 & 164\\
ALP & 122 & 118 & 123 & 124 & 123 & 119 & 118 & 119 & 122 & 113 & 118 & 153 & 118 & 136 & 120\\
Gorilla & \textbf{682} & 593 & 526 & 612 & 617 & 626 & 640 & 529 & \textbf{645} & \textbf{670} & \textbf{673} & \textbf{690} & \textbf{530} & \textbf{674} & \textbf{608}\\
Chimp & 495 & 456 & 505 & 567 & 516 & 498 & 509 & 478 & 553 & 482 & 437 & 611 & 507 & 577 & 498\\
DeXOR & 80 & 114 & 105 & 156 & 87 & 96 & 83 & 62 & 52 & 24 & 35 & 27 & 41 & 11 & 57\\
\midrule
\multicolumn{16}{@{}l}{\itshape Decode MB/s}\\
\vex{} \upper{} & 2{,}359 & 3{,}247 & 3{,}484 & 2{,}480 & 2{,}372 & 2{,}535 & 2{,}418 & 3{,}812 & 1{,}814 & 1{,}377 & 1{,}309 & 2{,}282 & 1{,}851 & 1{,}806 & 2{,}184\\
\vex{} \fast{} & \textbf{6{,}085} & \textbf{6{,}195} & \textbf{7{,}029} & \textbf{6{,}123} & \textbf{6{,}539} & \textbf{7{,}009} & \textbf{6{,}792} & \textbf{6{,}992} & \textbf{5{,}439} & 3{,}119 & 2{,}420 & \textbf{6{,}349} & 3{,}682 & 3{,}611 & \textbf{4{,}987}\\
Falcon & 179 & 198 & 209 & 185 & 187 & 193 & 185 & 176 & 183 & 127 & 131 & 38 & 147 & 49 & 165\\
ELF* & 633 & 548 & 501 & 590 & 590 & 554 & 483 & 589 & 605 & 414 & 453 & 562 & 458 & 916 & 528\\
ELF & 667 & 550 & 513 & 629 & 615 & 559 & 525 & 599 & 751 & 485 & 536 & 675 & 499 & 860 & 570\\
ALP & 4{,}603 & 4{,}614 & 4{,}692 & 4{,}596 & 4{,}594 & 4{,}605 & 4{,}591 & 4{,}605 & 4{,}487 & \textbf{4{,}345} & \textbf{4{,}352} & 4{,}511 & \textbf{4{,}483} & \textbf{4{,}129} & 4{,}542\\
Gorilla & 841 & 723 & 791 & 831 & 809 & 766 & 781 & 813 & 802 & 801 & 817 & 834 & 792 & 851 & 796\\
Chimp & 517 & 542 & 516 & 629 & 537 & 544 & 537 & 504 & 651 & 500 & 501 & 573 & 527 & 559 & 538\\
DeXOR & 369 & 369 & 370 & 504 & 326 & 353 & 299 & 333 & 310 & 192 & 210 & 241 & 248 & 160 & 301\\
\bottomrule
\end{tabular}
\renewcommand{\arraystretch}{1.0}
\vspace{2pt}\parbox{\textwidth}{\footnotesize CT=City-temp, IR=IR-bio-temp,
WS=Wind-Speed, P10=PM10-dust, SUK/USA/SDE=Stocks-UK/USA/DE,
DT=Dew-point-temp, AP=Air-pressure, BW=Basel-wind, BT=Basel-temp,
BP=Bitcoin-price, BM=Bird-migration, AS=Air-sensor.}
\end{table*}

\subsection{Exact Pages (Q1)}
\label{sec:eval-overall}

\textbf{Ratio.}  Table~\ref{tab:canonical} places both profiles and the
baselines on the same pages.  \upper{} reaches 7.53$\times$, 27.6\% fewer
bytes than Falcon, the strongest exact baseline, and 36.0\% fewer than Pcodec;
\fast{} reaches 7.01$\times$, 22.2\% fewer than Falcon and 40.2\% fewer than
ALP.  \upper{} is the smallest exact output on 13 streams, and \fast{} is
smaller than Falcon and ALP on every stream.  General-purpose compressors on
the same raw pages reach 3.71$\times$ (xz, preset~6) and 3.33$\times$ (zstd,
level~19), so \upper{} needs less than half of xz's bytes.  On Basel-temp, pages carry 3.92 minimal scales: \upper{}
reaches 5.61$\times$ against Falcon's 2.16$\times$, which keeps one scale per
chunk.  Basel-wind carries 8.15 scales per page, and \upper{} uses all four
lanes to reach 2.57$\times$ against Falcon's 2.09$\times$.  On Bird-migration, rational lattices hold the interpolated positions
(5.95$\times$ against 4.03$\times$), and on the stock streams, whose prices
mix tick sizes, exact gcds and split lattices give 12.8--13.8$\times$ against
8.9--12.3$\times$.  Air-sensor carries about 16 significant digits; there,
every codec stays between 1.22$\times$ and 1.38$\times$, and the binary lane
keeps \vex{} within 1\% of the best.

\textbf{Speed.}  \fast{} encodes at 476~MB/s, 4.0$\times$ ALP, and decodes at
4,987~MB/s, 1.10$\times$ ALP, with no entropy decoding.  It wins decode on
ten streams; ALP's unpacking, which carries no prefix dependency, wins on the
four where \fast{} routes several lanes, a rational lattice, or a binary lane.
\upper{} spends 6.1$\times$ the encode time of \fast{} on exact pricing,
encodes at 78~MB/s, 1.9$\times$ Falcon and 65\% of ALP, and decodes at
2,184~MB/s, four times ELF* (528~MB/s).  Gorilla and Chimp encode at 608 and
498~MB/s but stay below 1.46$\times$.  Figure~\ref{fig:tradeoff} places every
codec on both trade-offs.  The two profiles are the only points above
5.5$\times$, \upper{} at the ratio end and \fast{} at the speed end; \fast{}
dominates every baseline on decode, and on encode every baseline except
Gorilla and Chimp, which run 1.05--1.28$\times$ faster at a fifth of its
ratio.

\begin{table}[t]
\centering
\caption{\vex{} page size on the canonical streams: ratio and one-thread
encode / decode MB/s.}
\label{tab:page}
\footnotesize
\setlength{\tabcolsep}{3.0pt}
\begin{tabular}{@{}rrrrrrr@{}}
\toprule
 & \multicolumn{3}{c}{\fast{}} & \multicolumn{3}{c}{\upper{}}\\
\cmidrule(lr){2-4}\cmidrule(lr){5-7}
Page & Ratio & Enc. & Dec. & Ratio & Enc. & Dec.\\
\midrule
256 & 6.87 & 303 & 4{,}672 & 7.12 & 24 & 3{,}229\\
512 & 6.96 & 395 & 4{,}867 & 7.33 & 60 & 2{,}951\\
1{,}024 & 7.01 & 476 & 4{,}987 & 7.53 & 78 & 2{,}184\\
2{,}048 & 6.94 & 542 & 5{,}115 & 7.62 & 98 & 2{,}246\\
\bottomrule
\end{tabular}
\end{table}

\textbf{Page size.}  Table~\ref{tab:page} varies the page.  \fast{} stays
within 2.1\% of its best ratio from 256 to 2,048 values, since an 8-value
block carries only its width.  \upper{} gains 7\% from 256 to 2,048 values as
rANS tables amortize: it codes 8\% of its streams with rANS at 256 values and
76\% at 2,048, which is also why its decode rate is highest on short pages.
Its planning cost is largely per page, so its encode rate falls to 24~MB/s at
256 values, while \fast{} stays above 300~MB/s.  At every size \fast{} keeps
91--96\% of \upper{}'s ratio.

\textbf{Larger streams.}  On the 48-stream suite,
\upper{} reaches 9.61$\times$, 27.2\% fewer bytes than Falcon, and is the
smallest output on all 48 streams; \fast{} reaches 7.92$\times$, smaller than
Falcon and ALP on all 48, at 4.0$\times$ ALP's encode and 1.72$\times$ its
decode rate.  The building sensors store means and interpolations printed with
15 significant digits; rational lattices code them on their steps, so \upper{}
reaches 9.77$\times$ there against 1.73$\times$ for Falcon and 6.46$\times$ for
Pcodec, whose multiplier mode targets the same structure.  \upper{} encodes
these pages at 19~MB/s, the slowest domain, because every lane prices rational
and second-difference variants.

\subsection{Storage Engine (Q2)}
\label{sec:eval-tsfile}

\begin{table}[t]
\centering
\caption{Apache TsFile, one writer and one reader thread.  Top: four streams
(6.6~M values), ratio of complete files, MB/s.  Range reads the middle tenth
of each series by timestamp.  Camel is the patched Java codec in one warm
process, so its rates include that call.  Bottom: the 48 streams written as
785 flushes and merged per stream; Written is flushed plus merged bytes per
raw byte.}
\label{tab:tsfile}
\footnotesize
\setlength{\tabcolsep}{1.8pt}
\begin{tabular}{@{}lrrrrrr@{}}
\toprule
Value encoding & Ratio & +LZ4 & Write & Scan & Range & Page dec.\\
\midrule
\vex{} \upper{} & \textbf{6.96} & \textbf{7.76} & 48 & 210 & 194 & 2{,}248\\
\vex{} \fast{} & 6.27 & 6.93 & \textbf{94} & \textbf{226} & \textbf{200} & 6{,}348\\
DeXOR & 4.42 & 4.73 & 41 & 130 & 119 & 343\\
ELF* & 4.41 & 4.72 & 58 & 156 & 142 & 476\\
ELF & 3.79 & 4.04 & 69 & 166 & 148 & 619\\
ALP & 3.66 & 4.06 & 76 & 222 & 189 & \textbf{9{,}928}\\
Chimp & 1.53 & 1.64 & 76 & 119 & 100 & 266\\
Gorilla & 1.38 & -- & 57 & 93 & 83 & --\\
Camel & 1.12 & 1.40 & 12 & 29 & 30 & 33\\
\midrule
Flush$\to$merge & Flush & Merged & Written & Merge & Scan & \\
\midrule
\fast{}$\to$\upper{} & \textbf{6.33} & \textbf{7.39} & \textbf{0.29} & 36 & 125 & \\
\fast{}$\to$\fast{} & \textbf{6.33} & 6.35 & 0.32 & \textbf{56} & \textbf{131} & \\
ALP$\to$ALP & 3.59 & 3.60 & 0.56 & 48 & 129 & \\
Gorilla$\to$Gorilla & 1.69 & 1.69 & 1.19 & 35 & 80 & \\
\bottomrule
\end{tabular}
\end{table}

We register \vex{} as a TsFile value encoding, with one page format and the
profile chosen per writer (Table~\ref{tab:tsfile}).  Ratios count complete
files, so timestamps, headers, statistics, and indexes are charged to every
encoding; plain values make up 97.0\% of a plain file.  The \fast{} file takes
58.4\% of ALP's bytes and the \upper{} file 52.6\%, and \fast{} writes 24\%
faster than ALP.  Full scans run at ALP's rate, and so does a timestamp
range over the middle tenth of each series: \fast{} returns it at 200~MB/s
and ALP at 189~MB/s.  The reader spends most of its time outside the value
decoder, so \fast{}'s page decoder at 6.3~GB/s and ALP's at 9.9~GB/s leave
both queries bound by the file format.  Where LZ4 is on, it lifts the C++ encodings by 6--12\% and Camel by 26\%.
ALP+LZ4 reaches 65\% of \fast{}'s ratio without LZ4.

\textbf{Flush and compaction.}  We write the 48 streams as flushes of 96
pages, under IoTDB's 100,000-point flush threshold, and merge each stream's
flushes into one file, checking every value and timestamp.  Ingesting with
\fast{} and compacting with \upper{} stores the suite at 7.39$\times$, 48.7\%
of ALP's merged bytes, and writes 0.29 bytes per raw byte over both steps,
against 0.56 for ALP.  The merged \upper{} file is no larger than the \fast{}
file on any of the 48 streams, and the \upper{} merge runs at 36~MB/s per
core, 75\% of an ALP merge.

\textbf{Choosing a profile.}  These results suggest a simple policy.  \fast{}
suits ingestion: it encodes at about 0.5~GB/s per core, decodes fastest, and
emits only 7.4\% more bytes than \upper{} on the canonical pages.  \upper{}
suits compaction and cold data, where its encode cost is paid once and its
bytes persist.  Both write one format that one decoder reads, so the choice is
made per page and can change over a file's life.

\subsection{Mixed Resolutions (Q3)}
\label{sec:eval-mixed}

\begin{figure}[t]
\centering
\includegraphics[width=\columnwidth]{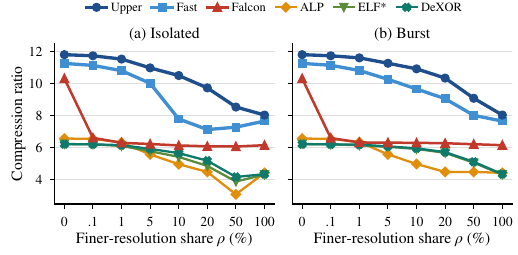}
\caption{Mixed-resolution sweep on the canonical pages: a share $\rho$ of
each page is written at scale~6 and the rest at scale~2, (a) spread evenly or
(b) as one run.}
\label{fig:mixed}
\end{figure}

The sweep rewrites each canonical page through a scale-2 decimal path and
writes a share $\rho$ of its positions at scale~6, with three seeds per
interior rate (Figure~\ref{fig:mixed}).  The curves follow the costs of
Section~\ref{sec:lanes}.  Falcon pays the lift at once: it falls from
10.31$\times$ to 6.62$\times$ at $\rho=0.1\%$, one fine value per page, and
stays near 6.2$\times$, the four-digit penalty of
Proposition~\ref{prop:penalty} paid on every step.  ALP pays per exception:
it loses 15\% at $\rho=5\%$ and reaches 3.10$\times$ at 50\%.  \upper{} moves
from 11.80$\times$ to 11.73$\times$ at 0.1\% and to 8.54$\times$ at 50\%, and
keeps 9.08$\times$ at 50\% when the fine values come in one run, which fills
one lane.  \fast{} tracks \upper{} up to 5\%; beyond the sparse limit of
$m/16$, the route of isolated fine values moves to 8-value blocks at about
two bits per value, and \fast{} holds 7.1--7.8$\times$.  Both profiles stay
above every baseline at every rate.

\subsection{Error-Bounded Pages (Q4)}
\label{sec:eval-lossy}

\begin{figure}[t]
\centering
\includegraphics[width=\columnwidth]{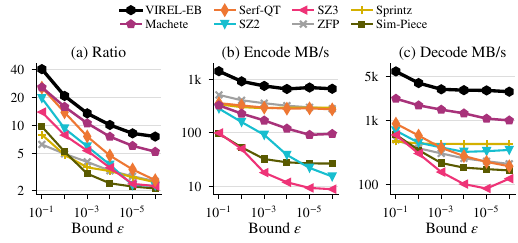}
\caption{Error-bounded codecs over 15 streams and six absolute bounds:
(a)~aggregate ratio, (b)~encode and (c)~decode MB/s.}
\label{fig:eb}
\end{figure}

\begin{table*}[t]
\centering
\caption{Bounded streams at $\varepsilon=10^{-3}$, independent 1,024-value
pages: compression ratio and one-thread throughput (higher is better).  Bold
marks the best ratio and the fastest rate.}
\label{tab:eb-all}
\footnotesize
\setlength{\tabcolsep}{1.6pt}
\renewcommand{\arraystretch}{0.92}
\begin{tabular}{@{}l*{16}{r}@{}}
\toprule
 & AP & BT & BW & CD & CT & DT & IR & MT & P10 & SG & USA & TD & TLa & TLo & WS & All\\
\midrule
\multicolumn{17}{@{}l}{\itshape Compression ratio}\\
\veb{} & \textbf{34.03} & \textbf{7.27} & \textbf{12.57} & \textbf{9.98} & \textbf{9.47} & \textbf{7.27} & \textbf{19.19} & \textbf{10.79} & \textbf{22.14} & \textbf{19.06} & \textbf{15.77} & \textbf{19.29} & \textbf{17.63} & \textbf{17.65} & \textbf{10.70} & \textbf{13.53}\\
Machete & 25.21 & 5.13 & 9.00 & 9.09 & 6.43 & 5.07 & 15.10 & 9.27 & 17.39 & 12.56 & 13.67 & 16.19 & 15.68 & 15.69 & 8.72 & 10.59\\
Serf-QT & 30.07 & 3.70 & 3.42 & 7.67 & 3.45 & 3.71 & 9.70 & 8.10 & 14.26 & 13.36 & 8.03 & 18.91 & 15.76 & 15.78 & 5.50 & 7.63\\
SZ2 & 18.45 & 2.37 & 5.83 & 5.92 & 3.40 & 2.47 & 8.32 & 5.48 & 10.19 & 6.35 & 8.29 & 10.17 & 11.00 & 11.03 & 4.06 & 5.96\\
SZ3 & 14.09 & 2.15 & 5.00 & 5.11 & 3.68 & 2.57 & 7.09 & 4.33 & 8.62 & 5.24 & 6.90 & 7.99 & 9.02 & 9.13 & 4.32 & 5.37\\
ZFP & 4.32 & 3.53 & 3.29 & 3.87 & 3.18 & 3.26 & 4.53 & 3.80 & 6.42 & 5.31 & 3.74 & 4.12 & 4.09 & 4.52 & 4.43 & 4.06\\
Sim-Piece & 6.59 & 1.85 & 2.53 & 3.06 & 2.46 & 1.81 & 3.34 & 2.27 & 6.45 & 6.02 & 2.84 & 4.36 & 3.23 & 3.23 & 2.72 & 3.03\\
Sprintz & 5.11 & 3.11 & 2.58 & 3.52 & 2.68 & 2.65 & 3.58 & 3.84 & 4.33 & 5.88 & 3.52 & 5.09 & 3.53 & 3.53 & 3.32 & 3.54\\
\midrule
\multicolumn{17}{@{}l}{\itshape Encode MB/s}\\
\veb{} & \textbf{892} & \textbf{631} & \textbf{656} & \textbf{525} & \textbf{582} & \textbf{962} & \textbf{1{,}065} & \textbf{715} & \textbf{577} & \textbf{760} & \textbf{810} & \textbf{612} & \textbf{913} & \textbf{915} & \textbf{781} & \textbf{750}\\
Machete & 301 & 66 & 171 & 204 & 114 & 68 & 244 & 164 & 223 & 171 & 247 & 198 & 303 & 306 & 151 & 170\\
Serf-QT & 359 & 258 & 285 & 270 & 284 & 261 & 286 & 286 & 321 & 338 & 270 & 312 & 311 & 311 & 271 & 295\\
SZ2 & 259 & 50 & 58 & 153 & 20 & 45 & 149 & 138 & 89 & 137 & 172 & 186 & 225 & 224 & 101 & 91\\
SZ3 & 14 & 15 & 15 & 20 & 9 & 8 & 42 & 13 & 38 & 59 & 14 & 13 & 15 & 62 & 44 & 18\\
ZFP & 350 & 335 & 339 & 352 & 331 & 331 & 380 & 328 & 488 & 444 & 322 & 330 & 332 & 368 & 411 & 356\\
Sim-Piece & 61 & 23 & 27 & 27 & 29 & 24 & 31 & 27 & 61 & 69 & 28 & 48 & 32 & 32 & 27 & 32\\
Sprintz & 301 & 305 & 289 & 295 & 306 & 300 & 275 & 298 & 284 & 313 & 283 & 292 & 270 & 270 & 289 & 287\\
\midrule
\multicolumn{17}{@{}l}{\itshape Decode MB/s}\\
\veb{} & \textbf{3{,}597} & \textbf{3{,}950} & \textbf{2{,}497} & \textbf{2{,}524} & \textbf{2{,}485} & \textbf{4{,}719} & \textbf{4{,}325} & \textbf{4{,}257} & \textbf{2{,}708} & \textbf{7{,}035} & \textbf{2{,}666} & \textbf{2{,}724} & \textbf{2{,}592} & \textbf{2{,}595} & \textbf{4{,}677} & \textbf{3{,}115}\\
Machete & 1{,}969 & 925 & 1{,}503 & 1{,}780 & 1{,}256 & 886 & 1{,}669 & 1{,}559 & 1{,}621 & 1{,}484 & 1{,}647 & 1{,}634 & 1{,}765 & 1{,}771 & 1{,}456 & 1{,}494\\
Serf-QT & 970 & 228 & 221 & 349 & 230 & 230 & 437 & 356 & 580 & 576 & 357 & 653 & 512 & 512 & 291 & 376\\
SZ2 & 619 & 319 & 298 & 325 & 243 & 353 & 416 & 311 & 441 & 378 & 380 & 406 & 541 & 545 & 262 & 382\\
SZ3 & 100 & 425 & 86 & 132 & 142 & 716 & 356 & 102 & 298 & 344 & 99 & 86 & 100 & 387 & 373 & 157\\
ZFP & 347 & 257 & 259 & 303 & 275 & 240 & 330 & 286 & 460 & 361 & 282 & 318 & 311 & 333 & 332 & 308\\
Sim-Piece & 401 & 148 & 181 & 196 & 184 & 153 & 227 & 189 & 372 & 418 & 194 & 300 & 223 & 223 & 189 & 217\\
Sprintz & 441 & 436 & 434 & 429 & 429 & 431 & 428 & 434 & 437 & 436 & 425 & 449 & 428 & 432 & 434 & 433\\
\bottomrule
\end{tabular}
\renewcommand{\arraystretch}{1.0}
\vspace{2pt}\parbox{\textwidth}{\footnotesize AP=Air-pressure, BT=Basel-temp,
BW=Basel-wind, CD=Chengdu-traj, CT=City-temp, DT=Dew-point-temp,
IR=IR-bio-temp, MT=Motor-temp, P10=PM10-dust, SG=Smart-grid, USA=Stocks-USA,
TD=T-drive, TLa/TLo=Tsbs-iot latitude/longitude, WS=Wind-Speed.}
\end{table*}

\veb{} has the highest aggregate ratio at every bound
(Figure~\ref{fig:eb}), from 40.43$\times$ at $\varepsilon=0.1$ to
7.61$\times$ at $10^{-6}$.  Machete is second at every bound, 25.78$\times$
and 5.23$\times$ at those two ends.  Over the 90 cells the geometric means
against Machete are 1.56$\times$, 5.18$\times$, and 2.60$\times$ on ratio,
encode, and decode.
\veb{} encodes at 658--1,415~MB/s and decodes at 2.9--5.9~GB/s, 4.0--7.7$\times$
and 2.1--2.8$\times$ Machete's aggregate rates.
Table~\ref{tab:eb-all} is every stream at $\varepsilon=10^{-3}$: \veb{} has
the best ratio and the fastest encode and decode on all 15, 13.53$\times$
overall at 750~MB/s encode and 3,115~MB/s decode.  The ratio gap is widest
on Air-pressure, 34.03$\times$ against Machete's 25.21$\times$, and narrowest
on Chengdu-traj, 9.98$\times$ against 9.09$\times$.

\begin{table}[t]
\centering
\caption{Reconstruction error: median RMSE over the 15 streams, in units of
$\varepsilon$.  Bold marks the smallest.}
\label{tab:eb-quality}
\footnotesize
\setlength{\tabcolsep}{3.6pt}
\begin{tabular}{@{}lrrrrrr@{}}
\toprule
$\varepsilon$ & $10^{-1}$ & $10^{-2}$ & $10^{-3}$ & $10^{-4}$ & $10^{-5}$ & $10^{-6}$\\
\midrule
\veb{} & 0.56 & 0.58 & 0.57 & 0.45 & \textbf{0.10} & \textbf{0.00}\\
Machete & 0.56 & 0.58 & 0.57 & 0.58 & 0.58 & 0.57\\
Serf-QT & 0.58 & 0.58 & 0.58 & 0.58 & 0.58 & 0.58\\
SZ2 & 0.55 & 0.56 & 0.56 & 0.54 & 0.43 & 0.42\\
SZ3 & 0.55 & 0.55 & 0.55 & 0.54 & 0.46 & 0.13\\
ZFP & \textbf{0.10} & \textbf{0.13} & \textbf{0.16} & \textbf{0.10} & 0.13 & 0.16\\
Sim-Piece & 0.45 & 0.50 & 0.32 & 0.21 & 0.14 & 0.17\\
Sprintz & 0.58 & 0.58 & 0.58 & 0.58 & 0.58 & 0.58\\
\bottomrule
\end{tabular}
\end{table}

\textbf{Accuracy.}  Table~\ref{tab:eb-quality} is the median RMSE over the
15 streams.  The largest error \veb{} measures is $(1-2^{-40})\,\varepsilon$.
At $\varepsilon\ge10^{-4}$ the smallest median is ZFP's, 0.10--0.16\,$\varepsilon$,
while \veb{}, Machete, Serf-QT, and Sprintz spread the error across the cell,
near the $\varepsilon/\sqrt3$ of a uniform error.  At $10^{-5}$ and $10^{-6}$
\veb{} is the smallest, 0.10\,$\varepsilon$ and 0.00\,$\varepsilon$: because
$\delta$ equals $2\varepsilon$ to within $2^{-40}$, a stream whose own
resolution is a multiple of $2\varepsilon$ lies on lattice points, as in the
thermometer example of Section~\ref{sec:bounded}.  In 33 of the 90 cells its
RMSE is below $0.01\,\varepsilon$, and the gcd that makes those values exact
is the one the stream coder removes, so the same structure lowers both bytes
and error.  Machete, Serf-QT, and Sprintz stay at 0.56--0.58\,$\varepsilon$
at every bound.  ZFP's lower error is paid for in the ratio of
Table~\ref{tab:eb-all}: at $\varepsilon=10^{-3}$ it reaches 4.06$\times$
against 13.53$\times$.

\textbf{What the bound buys.}  The ratios below are aggregates of the 15
Serf streams in this section, not the 14 canonical streams on which
\upper{} reaches 7.53$\times$ (Table~\ref{tab:canonical}).  On these 15,
exact \upper{} pages reach 6.88$\times$.  That already exceeds Machete once
the bound is $10^{-5}$ or finer: Machete's aggregate is 6.05$\times$ at
$10^{-5}$ and 5.23$\times$ at $10^{-6}$, the latter the 1,024-value entry of
Table~\ref{tab:eb-page}.  Exact \upper{} is also the smaller output on 14 of
the 15 streams at those two bounds.  Most of the streams are recorded at a
resolution coarser than $2\varepsilon$, so the decimal lattice holds them
without loss.  \veb{} adds 11\% over those exact pages at $10^{-6}$, 47\% at
$10^{-4}$, and 97\% at $10^{-3}$.  Choosing, per stream, the smaller of the
exact page and \veb{} adds at most 1.2\%, so a bounded column can keep
\veb{} alone.

\begin{table}[t]
\centering
\caption{Page size: aggregate ratio of \veb{} / Machete.}
\label{tab:eb-page}
\footnotesize
\setlength{\tabcolsep}{4pt}
\begin{tabular}{@{}rrrr@{}}
\toprule
Page & $\varepsilon=10^{-1}$ & $\varepsilon=10^{-3}$ & $\varepsilon=10^{-6}$\\
\midrule
256 & 35.7 / 16.2 & 12.6 / 7.9 & 7.25 / 4.34\\
512 & 38.4 / 21.8 & 13.0 / 9.4 & 7.43 / 4.85\\
1{,}024 & 40.43 / 25.78 & 13.53 / 10.59 & 7.61 / 5.23\\
2{,}048 & 41.5 / 28.1 & 13.8 / 11.4 & 7.72 / 5.49\\
\bottomrule
\end{tabular}
\end{table}

\textbf{Page size.}  Table~\ref{tab:eb-page} lists aggregate ratios only.
Encode and decode at the same four page lengths were timed in that sweep,
on the same 90 dataset--bound cells.  Across those cells the encode rate
stays at least 1.6$\times$ Machete's and the decode rate at least 1.2$\times$.
At 512 values and above, \veb{} wins ratio, encode, and decode together in
every cell; at 256 values it does so in 89 of the 90.  The short page still
favors the ratio: at 256 values and $\varepsilon=10^{-1}$, \veb{} reaches
2.2$\times$ Machete, 35.7 against 16.2 in the table, because its per-page
models cost tens of bits while Machete's coder restarts its adaptation in
every packet.  The 1,024-value row is the page length used in the rest of
this section: 40.43 against 25.78 at $\varepsilon=10^{-1}$, 13.53 against
10.59 at $10^{-3}$, and 7.61 against 5.23 at $10^{-6}$.  Lengthening the page
from 1,024 to 2,048 values raises Machete from 25.78$\times$ to 28.1$\times$
at $\varepsilon=10^{-1}$, while \veb{} moves only from 40.43$\times$ to
41.5$\times$.

\subsection{Mechanisms, Scaling, and Robustness (Q5)}
\label{sec:eval-mechanism}

\begin{table}[t]
\centering
\caption{Ablation: extra output bytes when one mechanism is disabled, and the
largest per-stream ratio loss.  \vex{} on the canonical pages; \veb{} on the
15 bounded streams at $\varepsilon=10^{-6}$.}
\label{tab:ablation}
\footnotesize
\setlength{\tabcolsep}{2.4pt}
\begin{tabular}{@{}lrrl@{}}
\toprule
Disabled mechanism & \fast{} & \upper{} & Largest loss\\
\midrule
Lanes (one lane per page) & 8.0\% & 6.2\% & Basel-temp, 26\%\\
Rational lattices & 4.4\% & 3.2\% & Bird-migration, 26\%\\
Coarse/residue split & 2.2\% & 2.0\% & Basel-temp, 15\%\\
Exact gcd & 2.0\% & 0.3\% & Stocks-DE, 16\%\\
rANS representations & -- & 5.6\% & Air-pressure, 11\%\\
Dictionary rANS & -- & 1.2\% & Air-pressure, 8\%\\
Second differences & -- & 0.2\% & Basel-temp, 2\%\\
Binary lane & 0.2\% & 0.2\% & Air-sensor, 6\%\\
\midrule
 & \multicolumn{2}{c}{\veb{}} & \\
\midrule
Exact gcd & \multicolumn{2}{c}{11.6\%} & \\
Coarse/residue split & \multicolumn{2}{c}{10.1\%} & \\
rANS representations & \multicolumn{2}{c}{6.5\%} & 2.4$\times$ faster decode\\
Phase origin & \multicolumn{2}{c}{0.6\%} & 7.0\% at $\varepsilon=0.1$\\
\bottomrule
\end{tabular}
\end{table}

\textbf{Ablation.}  Table~\ref{tab:ablation} removes one mechanism at a time.
Lanes are the largest single factor for \vex{}, and their gain concentrates
on the streams with the most minimal scales per page, as
Proposition~\ref{prop:penalty} predicts; collapsing \fast{} to one lane nearly
doubles its encode speed for 8\% more bytes.  Rational lattices carry the
averaged streams, and the split and the exact gcd each recover about 2\%
overall and up to 16\% on single streams.  In \upper{}, rANS saves 5.6\% at a
2.2$\times$ lower decode rate, which is the trade the two profiles expose.
For \veb{}, the shared gcd and split matter most at fine bounds, and the phase
at coarse ones.

\begin{table}[t]
\centering
\caption{Where the bits go, in bits per value.  Headers include scales,
anchors, gcds, stream parameters, and exceptions; escapes are the verbatim low
bits of hybrid rANS; framing is byte padding and page lengths.}
\label{tab:bits}
\footnotesize
\setlength{\tabcolsep}{3.0pt}
\begin{tabular}{@{}lrrrrr@{}}
\toprule
 & Total & Headers & Body & Escapes & Framing\\
\midrule
\fast{}, canonical & 9.13 & 0.45 & 8.65 & -- & 0.03\\
\upper{}, canonical & 8.50 & 0.40 & 4.88 & 3.19 & 0.03\\
\fast{}, 48 streams & 8.09 & 0.42 & 7.65 & -- & 0.02\\
\upper{}, 48 streams & 6.66 & 0.41 & 4.11 & 2.12 & 0.02\\
\veb{}, $\varepsilon=10^{-1}$ & 1.58 & 0.15 & 1.21 & 0.20 & 0.02\\
\veb{}, $\varepsilon=10^{-3}$ & 4.73 & 0.24 & 3.86 & 0.61 & 0.03\\
\veb{}, $\varepsilon=10^{-6}$ & 8.41 & 0.31 & 4.71 & 3.37 & 0.03\\
\bottomrule
\end{tabular}
\end{table}

\textbf{Where the bits go.}  Table~\ref{tab:bits} splits the output.  Headers
cost 0.15--0.45 bits per value, 4--9\% of each total, and exceptions
0.002 bits.  rANS shrinks the body of \upper{} from \fast{}'s 8.65 to 4.88 bits
per value and writes 3.19 bits as escapes, the low bits of large steps below
their two leading bits.  Escapes form 32--40\% of \upper{}'s bits and of
\veb{}'s at $10^{-6}$, while headers and framing together stay below half a
bit per value.

\begin{figure}[t]
\centering
\includegraphics[width=\columnwidth]{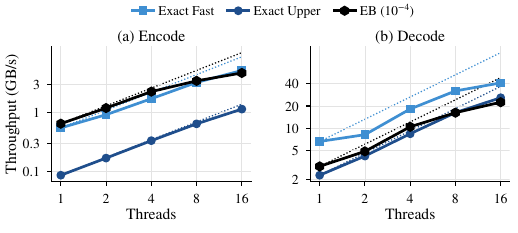}
\caption{Thread scaling with independent pages: \vex{} on the 48-stream
suite, \veb{} on its 15 streams at $\varepsilon=10^{-4}$.  Dotted lines show
linear scaling from one thread.}
\label{fig:scaling}
\end{figure}

\textbf{Scaling.}  Pages are independent, so both directions scale across
threads (Figure~\ref{fig:scaling}), and the output is byte-identical at every
thread count.  On 16 cores over the 48-stream suite, \fast{} encodes at
5.2~GB/s and decodes at 41~GB/s, and \upper{} encodes at 1.15~GB/s
(13.2$\times$ one thread) and decodes at 26~GB/s; \veb{} reaches 4.7~GB/s
encode and 22~GB/s decode.  The compute-bound \upper{} encoder scales almost
linearly, while the decoders approach memory bandwidth.

\textbf{Robustness.}  Randomized round trips draw signed zeros, NaN payloads,
infinities, subnormals, long fractions, coordinates near $2^{53}$, rational
grids, random page sizes, random disable masks, and both profiles, and compare
single- and multi-threaded output: 104,000 \vex{} runs with 247 million
values, 4,000 of them under AddressSanitizer and UBSan, and 32,000 \veb{}
runs, with no failure.

%% file: sections/08_related_work.tex
\section{Related Work}
\label{sec:related}

\textbf{Compression in data systems.}  Column stores choose encodings per block
and execute on compressed data \cite{abadi2006compression,btrblocks}.
Time-series systems add ordered ingestion, page and chunk boundaries, and
scan-heavy queries \cite{tsdbsurvey,survey,fptsurvey,hishida2025beyond}.
Gorilla brought XOR coding into an operational TSDB \cite{gorilla}, Apache
IoTDB exposes pluggable TsFile encodings per column \cite{iotdb,iotdbencoding},
and LSM engines choose different compressors for different levels
\cite{lsm,rocksdbcompression}.  \ravel{} keeps the page boundary, serves both
contracts behind one integer coder, and offers one encoder per storage path
over one format.

\textbf{Exact word codecs.}  FPC and FPZIP predict values or words and code the
residuals \cite{fpc,fpzip}.  Gorilla, TSXor, Chimp, Patas, ACTF, and Dolphin
refine XOR windows, references, zero coding, and alignment
\cite{gorilla,tsxor,chimp,patas,actf,dolphin}; Elf and Elf* erase a restorable
mantissa suffix before XOR coding \cite{elf,elfstar}, and AFC selects among
lossless strategies per block \cite{afc}.  These codecs model the binary word;
\vex{} models decimal coordinates and keeps words for its binary lane.

\textbf{Decimal codecs.}  ALP samples an exponent and factor per vector,
verifies each value by decoding it, packs the integers with frame of
reference, and stores the values that fail as exceptions \cite{alp}.  BtrBlocks' pseudodecimal encoding stores each double as digits
and a decimal exponent \cite{btrblocks}, and BUFF stores bounded-precision
values as byte-sliced fixed-point columns that queries scan in place
\cite{buff}.  Camel separates integer and fractional parts \cite{camel}, and
DeXOR realigns adjacent decimal prefixes and suffixes \cite{dexor}.  Ant and
Falcon convert values to integer surrogates before delta or bit-plane coding
\cite{ant,falcon}; Falcon proves when the conversion is recoverable and keeps
one decimal place per chunk.  Pcodec detects a common multiplier and
entropy-codes binned deltas \cite{pcodec}.  Where ALP checks one scale per
vector, \vex{} replays every value at its own minimal scale, with a sufficient
condition that covers more coordinates than a digit count; it gives each
resolution its own lane and history, replay-checks rational lattices per
lane, and plans the page by its emitted size.

\textbf{Error-bounded codecs.}  SZ predicts and quantizes with a grid of step
$2\varepsilon$ and entropy-codes the indices \cite{sz}, ZFP transforms blocks
\cite{zfp}, and MGARD and ISABELA target scientific fields
\cite{mgard,isabela}.  For time series, Sim-Piece fits piecewise-linear
segments \cite{simpiece}, Sprintz forecasts and bit-packs integers
\cite{sprintz}, Serf-QT quantizes XOR-style streams \cite{serf}, and Machete
combines Lorenzo prediction with a hybrid entropy coder \cite{machete}; see
\cite{ebsurvey} for a survey.  These codecs bound the error by analysis of
their quantizer; \veb{} checks each value with the decoder's own fused
multiply-add, chooses the lattice phase per page, and shares its integer
coder with the exact codec.

\textbf{Encoding selection.}  BtrBlocks picks a cascade of encodings per
block from a sample \cite{btrblocks}, ALP samples vectors to fix its exponent
and factor \cite{alp}, AFC selects a strategy per block \cite{afc}, and IoTDB
leaves the encoding to the schema \cite{iotdbencoding}.  These choices are
made by estimate before encoding.  \upper{} prices every candidate by the bits
it would emit and keeps the cheapest, and \fast{} ranks by a proxy but checks
its winner against the single lane exactly, so neither profile plans a page
larger than its single-lane page.

\textbf{Integer coding.}  Frame of reference, patched blocks, and vectorized
unpacking are standard integer tools \cite{pfor,dac}, rANS gives fast
table-driven entropy coding \cite{rans}.  Byte transforms and
general compressors \cite{bitshuffle,tdt,changebit,lz4,zstd,snappy} stack on
\ravel{} pages (Table~\ref{tab:tsfile}).  The \ravel{} stream coder combines
these tools with exact gcd factoring and a coarse/residue split, and prices
them per stream so that one decoder serves every encoder decision.

%% file: sections/09_conclusion.tex
\section{Conclusion}
\label{sec:conclusion}

We presented \ravel{}, route-local lattice residual compression that places a
value on a lattice only after replaying the decoder's arithmetic on it.
Decimal replay admits every exact value at its minimal scale, with a
closed-form guarantee below $2^{51}$, and fused multiply-add replay keeps
every bounded value within $\varepsilon$.  Lanes, rational
lattices, and a binary lane let one exact page hold several resolutions, and
one integer stream coder serves both contracts and two encoder profiles over
one format.  \upper{} emits 27.6\% fewer bytes than Falcon and is the
smallest output on all 48 larger streams; \fast{} beats ALP on ratio and both
speeds at once; ingest with \fast{} and compaction with \upper{} store the
48 streams in TsFile at 48.7\% of ALP's bytes; and \veb{} beats Machete on
ratio and both speeds in all 90 cells.
Replay makes the correctness argument the admission test.  One coder serves
both contracts, because a lane's gcd is also a bounded page's resolution.
Pricing candidates by the bits they emit sets the ratio ceiling.  Future work
covers multi-column lanes and relative error bounds.